\documentclass[10pt,twoside]{article}
\usepackage{latexsym}
\usepackage{amsmath}
\usepackage{amsfonts}
\usepackage{a4}
\def\volnum{}

\usepackage{aflbcours_enlarged}

\pdebut{1}
\def\pacs#1{\LP P.A.C.S.: #1}
\title{Schr{\"o}dinger's equation with gauge coupling derived from a 
continuity equation}
\titleshort{Schr{\"o}dinger's equation with gauge coupling \dots}
\author{Ulf Klein}
\authorshort{U. Klein}
\address{Johannes Kepler Universit{\"a}t Linz\\ 
              Institute for Theoretical Physics\\
              A-4040 Linz, Austria\\}

\begin{document}
\maketitle

\vskip 1cm

\bibliographystyle{plain}

\begin{abstract}
A quantization procedure without Hamiltonian is reported which starts
from a statistical ensemble of particles of mass $m$ and an 
associated continuity equation. The basic variables of this theory 
are a probability density $\rho$, and a scalar field $S$ which defines 
a probability current $\vec{j}=\rho\nabla S/m$. A first equation for $\rho$ and 
$S$ is given by the continuity equation. We further assume that this system 
may be described by a linear differential equation for a complex-valued 
state variable $\chi$. Using these assumptions and the simplest possible Ansatz 
$\chi(\rho,\,S)$, for the relation between $\chi$ and $\rho,\,S$, Schr\"odinger's 
equation for a particle of mass $m$ in a mechanical potential 
$V(q,t)$ is deduced. For simplicity the calculations are performed 
for a single spatial dimension (variable $q$). Using a second  
Ansatz $\chi(\rho,\,S,\,q,\,t)$, which allows for an explicit 
$q,t$-dependence of $\chi$, one obtains a generalized Schr\"odinger 
equation with an unusual external influence described by a 
time-dependent Planck constant. All other modifications of
Schr\"odinger' equation obtained within this Ansatz may be eliminated
by means of a gauge transformation. Thus, this second Ansatz 
may be considered as a generalized gauging procedure. Finally, 
making a third Ansatz, which allows for a \emph{non-unique} external 
$q,t$-dependence of $\chi$, one obtains Schr\"odinger's equation with 
electrodynamic potentials $\vec{A},\,\phi$ in the familiar gauge 
coupling form. This derivation shows a deep connection between 
non-uniqueness, quantum mechanics and the form of the gauge coupling. 
A possible source of the non-uniqueness is pointed out.
\end{abstract}

\section{Introduction}
\label{intro}
Usually a physical system, which one wants to describe 
quantum mechanically, is first identified in the context of 
classical physics and then somehow transferred to the quantum
mechanical domain; this process is referred to as "quantization". 
Often the first step in the quantization process is the tacit 
assumption that a Hilbert space is associated with the 
examined system. Then, the remaining task is to find the 
proper algebra of operators. A more direct method which avoids this 
assumption, is to ``derive'' Schr\"odinger's equation, i.e. to 
find premises which imply Schr\"odinger's equation. This may be done in several 
ways. The method which was historically at the beginning of quantum 
mechanics~\cite{schrodinger:quantisierung_I} ("wave mechanics") starts 
from the Hamilton-Jacobi equation of a classical system and tries to 
deduce from it - with the help of suitable 
modifications~\cite{cook:probability,castro.dutra:quantum,elizalde:quantum-hamilton-jacobi} - the corresponding quantum-mechanical equation for the 
time development of the system. Other premises leading to Schr\"odinger's 
equation include special assumptions about the structure of momentum 
fluctuations~\cite{hall.reginatto:schroedinger} and the principle of 
minimum Fisher information~\cite{reginatto:derivation}.

In this paper, a new quantization procedure is reported which shares with 
the last two examples the property that it does not start from a 
single-particle picture but from a statistical ensemble. The simplest 
nontrivial system, a spinless particle of mass $m$ in nonrelativistic 
approximation is investigated. In order to define the subject of this 
work more precisely we start from the classical Hamilton-Jacobi 
equation for the action function $S(\vec{q},t)$, which depends on 
the particle coordinates $q_k$ and the time $t$. It is given by     
\begin{equation}
  \label{eq:HJCL}
\frac{\partial S(\vec{q},t)}{\partial t}+\frac{1}{2m}
\left(\frac{\partial S(\vec{q},t)}{\partial\vec{q}}  \right)^{2}
+V(\vec{q},t)=0
\mbox{,}
\end{equation}
if the movement takes place under the influence of a potential 
$V(\vec{q},t)$. The momentum field, that appears in Eq.~(\ref{eq:HJCL})
is given by
\begin{equation}
  \label{eq:momenta}
p_k(\vec{q},t)=\frac{\partial S(\vec{q},t)}{\partial q_k}
\mbox{.}
\end{equation}

The fact that the Hamilton-Jacobi equation is the ideal starting 
point for the transition from classical physics to quantum mechanics 
is formally based on the following well-known reformulation of the 
time-dependent Schr\"odinger equation 
\begin{equation}
  \label{eq:TDSCHR}
\frac{\hbar}{\imath}\frac{\partial }{\partial t}\psi(\vec{q},t) + 
V(\vec{q},t) \psi(\vec{q},t) =
\frac{\hbar^2}{2m}\nabla\psi(\vec{q},t)
\mbox{.}
\end{equation}
If the complex-valued variable $\psi(\vec{q},t)$ is, without any 
restrictions of generality, written in the form 
\begin{equation}
  \label{eq:PRODANS}
\psi(\vec{q},t)=\sqrt{\rho(\vec{q},t)}\mathrm{e}^{\frac{\imath}{\hbar}S(\vec{q},t)}
\mbox{,}
\end{equation}
then one obtains from Eq.~(\ref{eq:TDSCHR}), by calculating the 
real parts of both sides, the relation
\begin{equation}
  \label{eq:CONT}
\frac{\partial \rho(\vec{q},t)}{\partial t}+\frac{\partial}{\partial\vec{q}}
\frac{\rho}{m} \frac{\partial S(\vec{q},t)}{\partial\vec{q}}=0
\mbox{,}
\end{equation}
which is a classical (no $\hbar$ occurs) 
continuity equation for the probability density 
$\rho$ and the probability current $\rho\,\vec{p}/m$.
Equating the imaginary parts of both sides of 
Eq.~(\ref{eq:TDSCHR}) one obtains the relation 
\begin{equation}
  \label{eq:QHJ}
\frac{\partial S(\vec{q},t)}{\partial t}+\frac{1}{2m}
\left( \frac{\partial S(\vec{q},t)}{\partial\vec{q}} \right)^{2}+V(\vec{q},t)=
\frac{\hbar^2}{2m}\frac{\triangle\sqrt{\rho(\vec{q},t)}}{\sqrt{\rho(\vec{q},t)}}
\mbox{,}
\end{equation}
which differs from the Hamilton-Jacobi equation~(\ref{eq:HJCL})
only by the single term on the right hand side. Eq.~(\ref{eq:QHJ}) is 
sometimes referred to as quantum Hamilton-Jacobi equation. 

Due to this similarity there have been 
attempts~\cite{castro.dutra:quantum,elizalde:quantum-hamilton-jacobi} 
to use Eq.~(\ref{eq:HJCL}) as a starting point and to derive 
Eq.~(\ref{eq:TDSCHR}), which presents the basis of quantum mechanics, 
by justifying introduction of the crucial quantum term appearing 
in Eq.~(\ref{eq:QHJ}). In the present work we go a different route, 
starting from assumptions which are simpler in certain respects. We 
postulate the existence of a statistical ensemble but do \emph{not} start 
from the Hamilton-Jacobi equation itself. Instead, our basic postulate is
the validity of a continuity equation (of the above type), interpreted
as a local conservation law of probability. Since we have two unknown 
functions ($\rho$ und $S$) and only one single equation, we clearly need 
further assumptions - and a second differential equation - in order 
to arrive at a mathematically well-defined problem. Our second 
assumption is very simple and of a purely formal nature. We require 
that both equations - the one already known and the second one still to 
be found - may be expressed mathematically as a single equation for 
a single complex state variable $\chi$. This second assumption expresses 
something like the postulate of maximal mathematical simplicity. As we 
know, this postulate may be quite successful in physics, in particular if 
combined with other ideas.

Thus, the continuity equation has to be ``extended'' to the complex 
domain. This task may be described briefly as follows: Consider a 
complex-valued variable $\chi$ which depends in an unspecified way on 
the real variables $S$ and $\rho$. Which differential equations for $\chi$ 
exist, whose real (or imaginary) part agrees with the continuity equation 
and which functional dependencies $\chi(\rho,S)$ are compatible with this 
requirement ? Note that both the functional dependence of $\chi$ and the 
shape of the differential equation are unknown ``variables'' of this 
problem. A detailed formulation of this problem is reported in the next 
section~\ref{sec:2}). The calculation, reported in section~\ref{sec:3}) 
and appendix~\ref{sec:8}, has been performed for simplicity for a single 
spatial dimension and for a reduced class of differential equations 
obeying several additional constraints. The result is Schr\"odinger's 
equation for a particle in an external mechanical potential. Further, 
we formulate and justify in section~\ref{sec:3}) the conjecture that 
all additional constraints except linearity may be omitted. 

In section~\ref{sec:4}) the original Ansatz is extended by allowing for 
an additional, explicit space-time dependence of the state variable
$\chi$. This is our first attempt to derive the minimal coupling rule, 
which is obeyed by essentially all fundamental interactions, in the 
present context. It turns out (details of the calculation are 
reported in appendix~\ref{sec:9}) that a second, very unusual external 
influence, besides the potential $V$, appears in the Schr\"odinger 
equation. It takes the form of a time-dependent Planck constant. All 
other modifications due to the extended Ansatz are spurious, because 
they may be eliminated from the Schr\"odinger equation by a gauge 
transformation. The gauge field itself cannot be derived by means of 
this Ansatz. In section~\ref{sec:5}) our second attempt is undertaken 
to derive a gauge field. The Ansatz of the last section is once more 
extended by allowing for a \emph{non-unique} space-time dependence 
of $\chi$. The result is Schr\"odinger's equation for a charged 
particle in an external electromagnetic field.

Section ~\ref{sec:6}) contains a detailed discussion of all
assumptions and results and may be consulted in a first reading 
to obtain an overview of this work. It also contains remarks of 
a speculative nature concerning the relation between the classical 
theory of charged particles and fields on the one hand and the 
form of quantum mechanics and gauge coupling on the other hand. 
In the last section~\ref{sec:7}) one finds concluding remarks.

\section{Formulation}
\label{sec:2}
We consider a statistical ensemble which is described by 
a probability density $\rho(\vec{q},t)$ and a probability current 
$\vec{j}=\rho\vec{p}/m$. We assume that the momentum field $\vec{p}$ may 
be written as the gradient of a scalar function $S(\vec{q},t)$, as 
in~(\ref{eq:momenta}); this simplest possible form of $\vec{p}$ holds 
in Hamilton-Jacobi theory~\cite{schiller:quasiclassical}. Our ensemble 
can thus be described by two variables, the real fields $\rho$ and $S$. From 
the continuity equation we obtain immediately Eq.~(\ref{eq:CONT}) as a 
first differential equation for $\rho$ and $S$. The physical meaning 
of the function $S(\vec{q},t)$ is still unclear. Of course, in the 
classical limit of the theory to be constructed, $S(\vec{q},t)$ should 
agree whith the action function of classical mechanics. 

We are confronted with the problem of finding a second differential 
equation for $\rho$ and $S$. Two solutions of this problem may be found 
in nature, classical statistical mechanics and quantum theory. Let 
us compare these two solutions trying to learn something about 
the quantization process. The second equation is~(\ref{eq:HJCL}) 
in classical physics and~(\ref{eq:QHJ}) in quantum theory. One sees 
immediately that in quantum theory the additional term in~(\ref{eq:QHJ}) 
implies a dramatic change of the basic concepts. This term is not an 
externally controlled input parameter but describes a coupling between 
the identity of the examined object and the statistics. But let us put 
aside these physical aspects. Instead let us ask the more formal 
question in which of the two theories the task of formulating the 
basic equations has been solved in a \emph{simpler} way. This is certainly 
quantum theory, because the two differential equations~(\ref{eq:CONT}) 
and~(\ref{eq:QHJ}) can be combined to a single equation, the 
Schr\"odinger equation~(\ref{eq:TDSCHR}) which has on top of that a 
simple linear structure. A similar unification is not possible in 
classical physics (one finds~\cite{schiller:quasiclassical} a nonlinear 
equation containing both $\psi$ and $\psi^{\star}$). This situation suggests that 
this principle of simplicity may be used, turning the logical direction 
around, to derive quantum mechanics from classical mechanics or to 
use it at least as an essential part of the quantization process. 

This principle of simplicity implies that the two differential 
equations for $\rho$ and $S$, namely Eq.~(\ref{eq:CONT}) and the second 
one which is still to be found, may be combined into a single equation 
for a two-component variable $\chi$, which we assume to be a complex quantity. 
The variable $\chi$ replaces $\rho$ and $S$ and should of course be a function 
of $\rho$ and $S$. If we define $\chi$ in terms of its real and imaginary parts 
(which has the advantage of linearity in comparison with the polar 
representation) $\chi$ takes in the simplest case (more general relations  
will be considered in the next sections) the form
\begin{equation}
  \label{eq:SHAPECHI}
\chi(\rho,S)=\chi_1(\rho,S) + \imath \chi_2(\rho,S)
\mbox{.}
\end{equation}
We assume furthermore that the differential equation for $\chi$, which 
will be referred to as ``state equation'', is a partial differential 
equation (with complex coefficients). The independent variables of 
this equation must be the usual space-time variables $q_k,\,t$, 
which also occur in the continuity equation. 

Each differential equation for $\chi$ whose real (or imaginary) part 
agrees with the continuity equation represents a possible 
extension of the latter. Its imaginary (or real) part provides us 
automatically with a second differential equation for our two 
variables. Thus, each one of these equations defines, in a purely  
formal sense, a physical theory motivated by the principle of 
simplicity. We denote the class of differential equations defined in
this way by $\mathbb{C}$. We know that $\mathbb{C}$ is not empty. It 
contains certainly Schr\"odinger's equation~(\ref{eq:TDSCHR}); it is a 
calculation of few lines to derive the continuity equation~(\ref{eq:CONT}) 
starting from~(\ref{eq:TDSCHR}). But the inverse problem, to start from 
the continuity equation~(\ref{eq:CONT}) and to derive the set of all 
extensions, i.e. the set $\mathbb{C}$, may be less trivial. In 
particular one cannot assume that $\mathbb{C}$ contains only a 
single equation.   

Being interested in the transition to quantum mechanics, we want to know 
under which conditions Schr\"odinger's equation can be derived. For each 
quantization procedure the important question to ask is  \emph{how} the 
final result is obtained. For that reason we make no use of 
symmetry considerations of any kind in this paper. Thus, which 
additional conditions are necessary in order to single out, from the set $\mathbb{C}$, Schr\"odinger's equation 
as only remaining equation ? We introduce for brevity a symbol, say 
$\mathbb{A}$, to denote the set of all additional conditions defined in
this way. $\mathbb{C}$ and $\mathbb{A}$ define together a quantization 
procedure. Obviously, this quantization procedure will only be convincing 
for a small number of (physically appealing) additional conditions, i. e. 
if $\mathbb{A}$ is ``small''. What we are looking for is the smallest set 
of such assumptions. 

If $\mathbb{A}$ should turn out to be empty, this would result in a 
very impressive quantization procedure; it would mean that  
quantum mechanics could be derived from only two assumptions, the
continuity equation and the existence of a complex state variable 
[as well as the tacit assumption that the laws of nature may be 
formulated as differential equations of the conventional type, 
compare section~\ref{sec:6})]. In the next section it will be
shown that this is not the case. However, the results indicate
that only a single additional condition, the linearity of the
differential equation, is required.   

How to find concretely the set $\mathbb{A}$ ? This set $\mathbb{A}$
is the smallest set of assumptions which eliminates all equations 
except the Schr\"odinger equation from the set $\mathbb{C}$. An obvious 
strategy is to start from a set of strong assumptions, say $\mathbb{A}_1$, 
which defines a small class of differential equations (all beeing loosely 
speaking ``similar'' to Schr\"odinger's equation). If the conditions implied by
$\mathbb{A}_1$ and $\mathbb{C}$ single out Schr\"odinger's equation, one may
eliminate some of the conditions contained in $\mathbb{A}_1$ and test if 
the corresponding, smaller set $\mathbb{A}_2$ of conditions may be 
used instead to lead to the same result. The final solution is the smallest 
possible set $\mathbb{A}$ obtained in this way.    

In this paper a complete solution to this problem is not 
given. All explicit calculations are performed using a 
set $\mathbb{A}_1$, defined by the following constraints
for the state equation:
\begin{itemize}
\item linearity,
\item a single space dimension (variable $q$),
\item only derivatives of first order in $t$ ,
\item only derivatives up to the second order in $q$,
\item no mixed derivatives. 
\end{itemize}   
However, the structure of the solutions leads to the  
conjecture on $\mathbb{A}$ mentioned above, which is 
discussed in more detail in the next section. 
\section{No interaction}
\label{sec:3}
In the framework of the strategy set up in the last section 
the possibility of interaction, i. e. of an external influence 
on our single particle system, was not taken into account. In 
principle there are two possibilities to do that; we may either modify 
the set $\mathbb{A}$ or the definition of $\chi$. In this work the second, 
more general, possibility will be chosen. In this sense, the simplest 
Ansatz $\chi=\chi(\rho,S)$, dealt with in the present section, describes a 
situation without interaction. The meaning of the term interaction 
defined in this way may differ (and does in fact differ) from the 
conventional one; we continue to use it for simplicity. 

We now have the following concrete mathematical problem: 
Find all differential equations obeying the conditions $\mathbb{A}_1$
whose real part (or imaginary part - this can be fixed arbitrarily  
and does not represent a real constraint) agrees - after  splitting off 
an arbitrary multiplicative factor - with the one-dimensional 
continuity equation. Our unknown variables are the coefficients of 
the differential equation, the state function $\chi$, and the 
factor $F$.

The basic complex-valued dynamic variable $\chi(\rho,S)$ is a function of 
the real variables $\rho(q,t)$ and $S(q,t)$ and is written in the 
above form~(\ref{eq:SHAPECHI}). The multiplicative factor may be 
an arbitrary complex function $F(\rho,S,q,t)$. The coefficients of 
the linear differential equation are denoted by $a,\,b,\,d,\,e$ 
and are arbitrary complex functions of $q,\,t$. Then, the fundamental 
requirement implied by $\mathbb{C}$ and $\mathbb{A}_1$ takes the form 
\begin{equation}
  \label{eq:FUNDEQ}
\Re\left[F\left(a\frac{\partial}{\partial t}\chi+b\frac{\partial}{\partial q}\chi+d\frac{\partial^{2}}{\partial q^{2}}\chi+
e\chi\right)\right]=\frac{\partial \rho}{\partial t}+
\frac{\partial \rho}{\partial q}\frac{1}{m}\frac{\partial S}{\partial q}+
\rho\frac{1}{m}\frac{\partial^{2}S}{\partial q^{2}}
\mbox{.}
\end{equation}
The right hand side of~(\ref{eq:FUNDEQ})
is given by the continuity equation. 

We have only a single equation~(\ref{eq:FUNDEQ})
and many unknowns, namely the complex functions 
$\chi(\rho,S),\,a(q,t),\,b(q,t),\,d(q,t),\,e(q,t),\,F(\rho,S,q,t)$.
However, Eq.~(\ref{eq:FUNDEQ}) is a very strong
requirement because the variables $\rho,S$ are 
arbitrary. Consequently, the coefficients of $\rho,S$ 
- and of all derivatives of $\rho,S$ - must agree on 
both sides of ~(\ref{eq:FUNDEQ}). Using this fact and 
introducing real and imaginary parts of $a,\ldots,e$ and 
$F$ according to $a=a_1+\imath a_2,\ldots,e=e_1+\imath e_2$
and $F=F_1+\imath F_2$, we obtain from Eq.~(\ref{eq:FUNDEQ}) 
the following 10 conditions   
\begin{eqnarray}
(a_1F_1-a_2F_2)\frac{\partial\chi_1}{\partial S}-(a_2F_1+a_1F_2)\frac{\partial\chi_2}{\partial S}&=& 0
\label{eq:BCOND1} 
\mbox{,} \\
(d_1F_1-d_2F_2)\frac{\partial^{2} \chi_1}{\partial \rho^{2}}-(d_2F_1+d_1F_2)
\frac{\partial^{2} \chi_2}{\partial \rho^{2}}&=& 0 \label{eq:BCOND2}
\mbox{,} \\
(d_1F_1-d_2F_2)\frac{\partial^{2} \chi_1}{\partial S^{2}}-(d_2F_1+d_1F_2)
\frac{\partial^{2} \chi_2}{\partial S^{2}}&=& 0 \label{eq:BCOND3}
\mbox{,} \\
(b_1F_1-b_2F_2)\frac{\partial\chi_1}{\partial \rho}-(b_2F_1+b_1F_2)\frac{\partial\chi_2}{\partial \rho}&=& 0
\label{eq:BCOND4A} 
\mbox{,} \\
(b_1F_1-b_2F_2)\frac{\partial\chi_1}{\partial S}-(b_2F_1+b_1F_2)\frac{\partial\chi_2}{\partial S}&=& 0
\label{eq:BCOND4B} 
\mbox{,} \\
(d_1F_1-d_2F_2)\frac{\partial\chi_1}{\partial \rho}-(d_2F_1+d_1F_2)\frac{\partial\chi_2}{\partial \rho}&=& 0
\label{eq:BCOND5} 
\mbox{,} \\
(a_1F_1-a_2F_2)\frac{\partial\chi_1}{\partial \rho}-(a_2F_1+a_1F_2)\frac{\partial\chi_2}{\partial \rho}&=& 1
\label{eq:BCOND6} 
\mbox{,} \\
(d_1F_1-d_2F_2)\frac{\partial^{2} \chi_1}{\partial \rho\partial S}-(d_2F_1+d_1F_2)
\frac{\partial^{2} \chi_2}{\partial \rho\partial S}&=&\,\frac{1}{2m}
\label{eq:BCOND7} 
\mbox{,} \\
(d_1F_1-d_2F_2)\frac{\partial\chi_1}{\partial S}-(d_2F_1+d_1F_2)\frac{\partial\chi_2}{\partial S}&=& 
\frac{\rho}{m}
\label{eq:BCOND8}
\mbox{,} \\ 
(e_1F_1-e_2F_2)\chi_1-(e_2F_1+e_1F_2)\chi_2&=& 0 
\label{eq:BCOND9} 
\mbox{.}
\end{eqnarray}  
We have 12 unknown quantities and 10 equations, but the unknown 
variables $a_1,a_2,\ldots,e_1,e_2$ and $F_1,F_2$ occur 
in~(\ref{eq:BCOND1})-~(\ref{eq:BCOND9}) only in the combinations
\begin{align}
&\bar{a}_1=a_1F_1-a_2F_2,\;\;\bar{a}_2=a_2F_1+a_1F_2\mbox{,}\label{eq:DEF1}\\
&\bar{b}_1=b_1F_1-b_2F_2,\;\;\bar{b}_2=b_2F_1+b_1F_2\mbox{,}\label{eq:DEF2}\\
&\bar{d}_1=d_1F_1-d_2F_2,\;\;\bar{d}_2=d_2F_1+d_1F_2\mbox{,}\label{eq:DEF3}\\
&\bar{e}_1=e_1F_1-e_2F_2,\;\;\bar{e}_2=e_2F_1+e_1F_2\label{eq:DEF4} 
\mbox{.}
\end{align}
Thus, we have 10 equations for 10 unknown quantities
$\bar{a}_1,\bar{a}_2,\ldots,\bar{e}_1,\bar{e}_2,\chi_1,\chi_2$.
Most of the equations~(\ref{eq:BCOND1})-~(\ref{eq:BCOND9})
look like differential equations. However, the coefficients 
apearing in these equations do also belong to our set of 
unknown variables. Thus, relations~(\ref{eq:BCOND1})-~(\ref{eq:BCOND9})
may either be used as differential equations for the unknown 
functions $\chi_1(\rho,S),\chi_2(\rho,S)\ldots$ or - if these
functions are already known - as constraints for the unknown 
coefficients. In the latter case, one has relations which must 
hold for arbitrary $\rho,\,S$, i.e. for each one of the 
considered conditions [out of the 
set~(\ref{eq:BCOND1})-~(\ref{eq:BCOND9})]
the coefficients of all linear independent functions 
of $\rho,\,S$ must vanish separately. While this type of problem 
may be somewhat unusual, its solution is straightforward and 
does not require any unusual mathematical methods (its solution 
could possibly be simplified by using more sophisticated methods). 

The first part of the calculation, the determination of 
$\chi_i,\,F_i$ is described in detail in appendix~\ref{sec:8}.
The result is  
\begin{eqnarray}
\chi_1(\rho,S)&=& -\frac{2m|d|^2}{c_2}
\sqrt{\rho}
\cos\left(\frac{c_2}{2m|d|^2}S+C_5 \right)
\mathrm{e}^{-\frac{c_1}{2m|d|^2}S-C_6}+C_3
  \label{eq:DRFXI1}
\mbox{,} \\
\chi_2(\rho,S)&=& -\frac{2m|d|^2}{c_2}
\sqrt{\rho}
\sin\left(\frac{c_2}{2m|d|^2}S+C_5 \right)
\mathrm{e}^{-\frac{c_1}{2m|d|^2}S-C_6}+C_4
\label{eq:DRFXI2}
\mbox{,} 
\end{eqnarray}
and
\begin{eqnarray}
F_1(\rho,S,q,t)&=& \frac{\sqrt{\rho}}{m|d|^2}
\bigg[ 
d_1\sin\left(\frac{c_2}{2m|d|^2}S+C_5 \right)
\nonumber\\
&+&d_2\cos\left(\frac{c_2}{2m|d|^2}S+C_5 \right)
 \bigg]
\mathrm{e}^{\frac{c_1}{2m|d|^2}S+C_6}
   \label{eq:DRFFF1}
\mbox{,} \\
F_2(\rho,S,q,t)&=&\frac{\sqrt{\rho}}{m|d|^2}
\bigg[ 
d_1\cos\left(\frac{c_2}{2m|d|^2}S+C_5 \right)
\nonumber\\
&-&d_2\sin\left(\frac{c_2}{2m|d|^2}S+C_5 \right)
 \bigg]
\mathrm{e}^{\frac{c_1}{2m|d|^2}S+C_6}
\label{eq:DRFFF2}
\mbox{,} 
\end{eqnarray}
where $c_1,\,c_2$ are linear combinations of coefficients, defined 
in Eq.~(\ref{eq:ABKKN}), and $|d|$ is the modulus of the complex number $d$.

Only a part of all conditions, namely~(\ref{eq:BCOND1}),
(\ref{eq:BCOND2}), (\ref{eq:BCOND5}), (\ref{eq:BCOND6}), 
(\ref{eq:BCOND7}), and (\ref{eq:BCOND8}) have been used in the 
course of the calculations leading to 
Eqs~(\ref{eq:DRFXI1})-(\ref{eq:DRFFF2}) (see appendix~\ref{sec:8}). 
In these conditions only the coefficients $a_i$ and $d_i$ appear. 
Conditions (\ref{eq:BCOND4A}),(\ref{eq:BCOND4B}), containing the $b_i$, 
and condition (\ref{eq:BCOND9}), containing the $e_i$, have not been 
used. The conditions not yet used play the role of constraints for
the constants of integration $C_3,\,C_4,\,C_5,\,C_6$ and the real-
and imaginary parts of the coefficients $a,\,b,\,d,\,e$. We will, for 
brevity, refer to the totality of all these quantities as ``parameters''. 

Our $12$ parameters are constants with respect to the variables
$\rho,\,S$, but may be arbitrary functions of $q,t$, as far as the 
calculation reported in appendix~\ref{sec:8} is concerned.    
However, they have to obey the basic requirement of this section,
that $\chi$ depends only on $\rho,\,S$, and not on $q,t$. This implies
that the following parameters or combinations of parameters 
\begin{equation}
  \label{eq:KFK7P}
C_3,\,C_4,\,C_5,\,C_6,\,\frac{c_1}{2m|d|^2},\,\frac{c_2}{2m|d|^2}
\mbox{}
\end{equation}
are constants (we use this term now in the usual sense of being 
independent of $q,t$). 
 
In order to obtain the explicit form of the constraints for the 
parameters, Eqs.~(\ref{eq:DRFXI1})-(\ref{eq:DRFFF2}) have to 
be inserted in the conditions not yet 
used~(\ref{eq:BCOND3}), (\ref{eq:BCOND4A}),(\ref{eq:BCOND4B}) 
and (\ref{eq:BCOND9}). These conditions hold for arbitrary values 
of $\rho,S$. This leads to $C_3=C_4=0$ while no constraints exist 
for $C_5,\,C_6$. For the coefficients one obtains the conditions
\begin{align}
&c_1=a_1 d_1+a_2 d_2=0\mbox{,}\label{eq:DFQ1}\\
&d_1 b_1+d_2 b_2=0\mbox{,}\label{eq:DFQ2}\\
&d_1 b_2-d_2 b_1=0\mbox{,}\label{eq:DFQ3}\\
&a_2 d_1-a_1 d_2=r_1 \left(d_1^2+d_2^2 \right)\mbox{,}\label{eq:DFQ4}\\
&e_2 d_1-e_1 d_2=0\mbox{,}\label{eq:DFQ5}
\end{align} 
where $r_1$ in~(\ref{eq:DFQ4}) is an arbitrary real constant.
As a consequence of~(\ref{eq:DFQ1})-(\ref{eq:DFQ5}) all parameters 
(with the exception of the arbitrary real constants $C_5$ und $C_6$) 
may be expressed in terms of three arbitrary real functions 
$d_1(q,t)$, $d_2(q,t)$, $f(q,t)$ and the real number $r_1$. 
In terms of the complex coefficients one obtains the following 
result: $d$ is an arbitrary complex-valued function of $q$ und 
$t$, $b=0$, and $a$ and $e$ are determined by $d$ according 
to the linear relation 
\begin{equation}
  \label{eq:RFDTK}
a=\imath r_1 d,\;\;\;e=f d
\mbox{.}
\end{equation}
Now, all conditions have been taken into account. 
The complex quantities $\chi$ and $F$ may we written as
\begin{eqnarray}
  \chi &=& -\frac{2m}{r_1}\mathrm{e}^{-C_6}\sqrt{\rho}
\mathrm{e}^{\imath \left(\frac{r_1}{2m}S+C_5 \right)}
   \label{eq:KMPCHI}
\mbox{,} \\
 F &=& \imath \frac{\sqrt{\rho}}{md}\mathrm{e}^{C_6}
\mathrm{e}^{-\imath \left(\frac{r_1}{2m}S+C_5 \right)}
   \label{eq:DASFC}
\mbox{,}
\end{eqnarray}
and the differential equation for $\chi$ takes the form
\begin{equation}
  \label{eq:DGLICF}
\imath r_1 d \frac{\partial\chi}{\partial t}+d\frac{\partial^{2}\chi}{\partial q^{2}}+f d \chi = 0
\mbox{.}
\end{equation}

If $S$ obeys the continuity equation~(\ref{eq:CONT}), then
it has the dimension of an action. Consequently, the constant 
$2m/r_1$ has the dimension of an action; we identify this constant 
with Planck's constant $\hbar$. The constant $d$ may be canceled 
and the constant $e$ has the dimension $cm^{-2}$. We may now come 
back to a more conventional notation by introducing quantities $\psi(q,t)$ 
and $V(q,t)$, which replace $\chi(q,t)$ and $f(q,t)$ and are defined by           
\begin{equation}
  \label{eq:DFNVZG}
\psi(q,t)=\sqrt{\rho(q,t)}\mathrm{e}^{\imath \frac{S(q,t)}{\hbar}},
\;\;\;\;V(q,t)=-\frac{\hbar^{2}}{2m}f(q,t)
\mbox{.}
\end{equation}
Then, Eq.~(\ref{eq:DGLICF}) takes the form 
\begin{equation}
  \label{eq:SCHRILY}
-\frac{\hbar}{\imath}\frac{\partial\psi}{\partial t}=
-\frac{\hbar^{2}}{2m}\frac{\partial^{2}\psi}{\partial q^{2}}+
V(q,t)\psi
\mbox{,}
\end{equation}
which agrees with the one-dimensional Schr\"odinger equation 
for a particle of mass $m$ in an external, mechanical potential 
$V(q,t)$ (the terms containing $C_5,\,C_6$, which correspond to 
the usual freedom in amplitude and phase, have not been written 
down). 

Thus, using the Ansatz $\chi=\chi(\rho,S)$ and the set of conditions $\mathbb{A}_1$
we found that only the one-dimensional Schr\"odinger equation yields the 
continuity equation as its real part. More precisely, we found an 
infinite number of such equations differing from each other by the choice 
of an arbitrary function $V(q,t)$. We obtained a coupling to an external 
potential but no coupling to a gauge field; the meaning of our term 
``no interaction'' should be adjusted accordingly. If the interaction 
is considered as a secondary aspect, the above assumptions define a 
quantization procedure, i.e. a way to perform the transition from 
classical physics to quantum mechanics. 
    
Can we improve this quantization procedure by reducing the number
of assumptions contained in $\mathbb{A}_1$ ? Let us keep first the 
linear structure as well as the one-dimensionality of the equation 
and allow for higher derivatives with respect to $t$ und $q$. Thus, 
new terms appear in the fundamental condition~(\ref{eq:FUNDEQ}). Will 
these new terms survive or will the corresponding coefficients vanish? 
All derivatives of $\rho$ and $S$, which occur in the continuity equation,
on the right hand side of~(\ref{eq:FUNDEQ}), appeared already in the 
differential equation defined by $\mathbb{A}_1$. Therefore, all the 
new terms, due to the higher derivatives, will have to vanish (the 
corresponding coefficients are all solutions of homoneneous equations) 
just as, in the above calculation, the term proportional to the first 
derivative with respect to $q$ had to vanish.
  
As a next step, let us allow for three spatial dimensions and keep only 
the condition of linearity. To get an idea what happens in this 
case, preliminary calculations using, instead of~(\ref{eq:FUNDEQ}), the 
fundamental condition  
\begin{equation}
  \label{eq:FUNDEQNEM}
\begin{split}
&\Re\left[F\left(a\frac{\partial}{\partial t}\chi+b_k\frac{\partial}{\partial q_k}\chi+
c_k \frac{\partial}{\partial t} \frac{\partial}{\partial q_k} \chi
+ d_{ik} \frac{\partial}{\partial q_i} \frac{\partial}{\partial q_k} \chi+
f\frac{\partial^{2}}{\partial t^{2}}\chi+ e\chi\right)\right]
=\\
&\frac{\partial \rho}{\partial t}+
\frac{\partial \rho}{\partial q_k}\frac{1}{m}\frac{\partial S}{\partial q_k}+
\rho\frac{1}{m}\frac{\partial}{\partial q_{k}}\frac{\partial}{\partial q_{k}}S
\end{split}
\mbox{,}
\end{equation}
have been performed (summation from $1$ to $3$ over double indices). 
The differential equation displayed in~(\ref{eq:FUNDEQNEM}) contains
all derivatives of $q_k,\,t$ up to second order taking into acount
also the possibility of anisotropic coefficients. These (not completed)
calculations indicate (i) that the anisotropy of the coefficients is, as 
one would expect intuitively, suppressed by the isotropy of the 
continuity equation, (ii) that the coefficients of the higher 
derivatives vanish, for the reason just mentioned, and (iii) that 
an exact calulation starting from~(\ref{eq:FUNDEQNEM}), 
which is completely analogous to the one above, leads to the 
three-dimensional Schr\"odinger equation~(\ref{eq:TDSCHR}). 

The remaining condition in $\mathbb{A}_1$ is the linearity of 
the differential equation. It can not be eliminated; one realizes
immediately that e.g. a cubic nonlinearity of the form
$r|\psi|^{2}\psi$, where $r$ is real, is compatible with the 
continuity equation. Thus we conjecture that our final 
quantization procedure is defined by a smallest set $\mathbb{A}$, 
which contains only a single condition, namely the linearity 
of the differential equation. 
 
What is the physical meaning of the remaining condition of linearity ?
A plausible explanation is that linearity is a consequence of the
probabilistic interpretation~\cite{home:ensemble} of the wave 
function, which excludes (in general) predictions about single events. 
There is a large literature on the relation between linearity and 
indeterminism; we only mention Mackey's axioms for statistical 
theories~\cite{mackey:mathematical} and Caticha's derivation of 
quantum mechanics from the rules of manipulating probability 
amplitudes~\cite{caticha:consistency}. On the other hand it is 
well-known that macroscopic quantum phenomena exist in nature, such 
as superconductivity and superfluidity, which may be successfully 
described by \emph{nonlinear} terms in the corresponding complex state 
variables. This is not in conflict with the above explanation. 
In contrast to the true wave function, the state variables 
of these many-body theories do not  possess a probabilistic 
interpretation. They lost their ``immaterial'' (probabilistic) 
meaning in the thermodynamic limit and may be directly measured 
in single experiments. The condition of linearity is required to 
exclude such theories from the consideration.

\section{First attempt to derive interaction}
\label{sec:4}
The Ansatz $\chi=\chi(\rho,\,S)$ used in the last section led, somewhat unexpectedly, 
to a term $V\psi$ in Schr\"odinger's equation~(\ref{eq:SCHRILY}), describing 
an interaction by means of an external mechanical Potential $V$. A term 
describing coupling to a gauge field did, not unexpectedly, not appear. In 
this section we start to study the following question: Is it possible to 
obtain this type of coupling, which is obeyed by all fundamental interactions, 
in the present framework ? If possible it requires, at any rate, a 
different, properly generalized Ansatz. 

There is an obvious possibility to generalize our Ansatz with regard to 
an external influence: we may allow for an additional explicit 
$q,\,t$-dependence of our state variable $\chi$, i.e. write  
\begin{equation}
  \label{eq:ERWANS1}
\chi=\chi(\rho,S,q,t)
\mbox{,}
\end{equation}
instead of $\chi=\chi(\rho,S)$. Such an extension seems quite natural if one 
wants to describe an external, otherwise unspecified influence on 
the system described by $\chi$. We will study the consequences 
of~(\ref{eq:ERWANS1}) keeping all other assumptions unchanged. 

Using~(\ref{eq:ERWANS1}) does not change the appearance of the 
basic condition~(\ref{eq:FUNDEQ}) in a fundamental way but 
additional derivatives with respect to the explicit $q,\,t$-dependence 
have to be taken into account. To indicate this difference, we replace 
the symbols for the partial derivatives by symbols for total derivatives. 
Thus, our fundamental condition, generalized according to the new 
Ansatz~(\ref{eq:ERWANS1}) takes the form 
\begin{equation}
  \label{eq:FUNDEQIA}
\Re\left[F\left(a\frac{\mathrm{d}}{\mathrm{d}t}\chi+
b\frac{\mathrm{d}}{\mathrm{d} q}\chi+
\mathrm{d}\frac{\mathrm{d}^{2}}{\mathrm{d}q^{2}}\chi+
e\chi\right)\right]=\frac{\partial \rho}{\partial t}+
\frac{\partial \rho}{\partial q}\frac{1}{m}\frac{\partial S}{\partial q}+
\rho\frac{1}{m}\frac{\partial^{2}S}{\partial q^{2}}
\mbox{.}
\end{equation}
The form of the factor $F(\rho,S,q,t)$ and the coefficients $a,b,d,e$
remain unchanged; the latter may be arbitrary functions of $q,t$. 
The problem defined by~(\ref{eq:FUNDEQIA}) contains now, in comparison 
to~(\ref{eq:FUNDEQ}), an even larger number of unknown functions. 
Fortunately, it will turn out that the equations to determine the 
$\rho,S$-dependence and those for the $q,t$-dependence are ``decoupled'' 
in the sense that they may be solved one after the other.  

Comparing the coefficients of the derivatives on both sides of 
Eq.~(\ref{eq:FUNDEQIA}) we now obtain the following 10 conditions 
\begin{eqnarray}
\bar{a_1}\frac{\partial\chi_1}{\partial S}-\bar{a_2}\frac{\partial\chi_2}{\partial S}&=& 0
\label{eq:BCOND1NE} 
\mbox{,} \\
\bar{d_1}\frac{\partial^{2} \chi_1}{\partial \rho^{2}}-\bar{d_2}
\frac{\partial^{2} \chi_2}{\partial \rho^{2}}&=& 0 \label{eq:BCOND2NE}
\mbox{,} \\
\bar{d_1}\frac{\partial^{2} \chi_1}{\partial S^{2}}-\bar{d_2}
\frac{\partial^{2} \chi_2}{\partial S^{2}}&=& 0 \label{eq:BCOND3NE}
\mbox{,} \\
\bar{b_1}\frac{\partial\chi_1}{\partial \rho}-\bar{b_2}\frac{\partial\chi_2}{\partial \rho}
+2\Big(\bar{d_1}\frac{\partial^{2}\chi_1}{\partial \rho\partial q} 
&-&\bar{d_2}\frac{\partial^{2}\chi_2}{\partial \rho\partial q} \Big)
= 0
\label{eq:BCOND4ANE} 
\mbox{,} \\
\bar{b_1}\frac{\partial\chi_1}{\partial S}-\bar{b_2}\frac{\partial\chi_2}{\partial S}
+2\Big(\bar{d_1}\frac{\partial^{2}\chi_1}{\partial S\partial q} 
&-&\bar{d_2}\frac{\partial^{2}\chi_2}{\partial S\partial q} \Big)
= 0
\label{eq:BCOND4BNE} 
\mbox{,} \\
\bar{d_1}\frac{\partial\chi_1}{\partial \rho}-\bar{d_2}\frac{\partial\chi_2}{\partial \rho}&=& 0
\label{eq:BCOND5NE} 
\mbox{,} \\
\bar{a_1}\frac{\partial\chi_1}{\partial \rho}-\bar{a_2}\frac{\partial\chi_2}{\partial \rho}&=& 1
\label{eq:BCOND6NE} 
\mbox{,} \\
\bar{d_1}\frac{\partial^{2} \chi_1}{\partial \rho\partial S}-\bar{d_2}
\frac{\partial^{2} \chi_2}{\partial \rho\partial S}&=&\,\frac{1}{2m}
\label{eq:BCOND7NE} 
\mbox{,} \\
\bar{d_1}\frac{\partial\chi_1}{\partial S}-\bar{d_2}\frac{\partial\chi_2}{\partial S}&=& 
\frac{\rho}{m}
\label{eq:BCOND8NE}
\mbox{,} \\ 
\bar{e_1}\chi_1-\bar{e_2}\chi_2
+\bar{a_1}\frac{\partial\chi_1}{\partial t}-\bar{a_2}\frac{\partial\chi_2}{\partial t}
+\bar{b_1}\frac{\partial\chi_1}{\partial q}&-&\bar{b_2}\frac{\partial\chi_2}{\partial q}
+\bar{d_1}\frac{\partial^{2} \chi_1}{\partial q^{2}}-\bar{d_2}\frac{\partial^{2} \chi_2}{\partial q^{2}}
= 0 
\label{eq:BCOND9NE} 
\mbox{.}
\end{eqnarray}
Again, each one of the conditions~(\ref{eq:BCOND1NE})-(\ref{eq:BCOND9NE}) 
may split into several sub-conditions, if on the left hand side 
several linear independent functions of $\rho,\,S$ occur. 
Comparing with the previous ``interaction-free'' 
conditions~(\ref{eq:BCOND1})-(\ref{eq:BCOND9}), we 
see that seven of the Eqs.~(\ref{eq:BCOND1NE})-(\ref{eq:BCOND9NE}) agree 
with corresponding equations in the set~(\ref{eq:BCOND1})-(\ref{eq:BCOND9}).
Only~(\ref{eq:BCOND4ANE}),~(\ref{eq:BCOND4BNE}), and~(\ref{eq:BCOND9NE}) 
differ from the corresponding previous 
conditions~(\ref{eq:BCOND4A}),~(\ref{eq:BCOND4B}),
and~(\ref{eq:BCOND9}) by new terms. These new terms appear as a 
consequence of the explicit $q,\,t$-dependence of $\chi_i$ and lead 
obviously to a coupling of previously independent coefficients. 

A great simplification of the problem defined 
by~(\ref{eq:BCOND1NE})-(\ref{eq:BCOND9NE}) takes place as a
consequence of the fact that 
conditions~(\ref{eq:BCOND4A}),~(\ref{eq:BCOND4B}) 
and~(\ref{eq:BCOND9}) have not been used during the calculation of 
the $\rho,\,S$-dependence of $\chi$ in section~\ref{sec:3}) (only conditions 
containing $a$ and $d$ were required). Therefore all conditions 
in the set~(\ref{eq:BCOND1NE})-(\ref{eq:BCOND9NE}) which are necessary 
to obtain this relationship remain unchanged. Since the additional 
$q,\,t$-dependence does not affect this part of the calculation, the 
formal results for $\chi$ and $F$ from section~\ref{sec:3}) may be taken 
over without modification for the present calculation. Thus, for the 
real and imaginary parts of $\chi$ we obtain
\begin{eqnarray}
\chi_1(\rho,S,q,t)&=& -\frac{2m|d|^2}{c_2}
\sqrt{\rho}
\cos\left(\frac{c_2}{2m|d|^2}S+C_5 \right)
\mathrm{e}^{-\frac{c_1}{2m|d|^2}S-C_6}+C_3
  \label{eq:DRFXI1NE}
\mbox{,} \\
\chi_2(\rho,S,q,t)&=& -\frac{2m|d|^2}{c_2}
\sqrt{\rho}
\sin\left(\frac{c_2}{2m|d|^2}S+C_5 \right)
\mathrm{e}^{-\frac{c_1}{2m|d|^2}S-C_6}+C_4
\label{eq:DRFXI2NE}
\mbox{.} 
\end{eqnarray}
The real and imaginary parts of $F(\rho,S,q,t)$ remain completely 
unchanged and are given by Eqs.~(\ref{eq:DRFFF1}) and~(\ref{eq:DRFFF2}).
Of course, there is an important difference between the results 
for $\chi$ and $F$ of the last and the present section. In 
Eqs.~(\ref{eq:DRFXI1NE}) and~(\ref{eq:DRFXI2NE}) not only 
the coefficients but also the integration constants 
$C_3,\,C_4,\,C_5,\,C_6$ may be arbitrary functions of $q,t$ 
(similar remarks apply to $F$). Those conditions, from the 
set~(\ref{eq:BCOND1NE})-(\ref{eq:BCOND9NE}), which have not 
yet been used present constraints for the spatial and temporal 
variation of all these parameters.  

In order to find the explicit form of these constraints the 
results~(\ref{eq:DRFXI1NE}),~(\ref{eq:DRFXI2NE}),~(\ref{eq:DRFFF1}), 
and ~(\ref{eq:DRFFF2}) have to be inserted in the 
conditions not yet used, 
namely~(\ref{eq:BCOND3NE}),~(\ref{eq:BCOND4ANE})~(\ref{eq:BCOND4BNE}) 
and~(\ref{eq:BCOND9NE}). The calculation reported in detail in 
appendix~\ref{sec:9} leads to the following results for the parameters.
The quantities $C_5,\,C_6$ are arbitrary functions of $q,\,t$, while 
$C_3=C_4=0$. The complex coefficient $d$ is an arbitrary function 
of $q,\,t$. The remaining coefficients are given by 
\begin{eqnarray}
a&=&\imath 2 m \tilde{u} d  
\label{eq:FJKUZA}
\mbox{,} \\
b&=& \left(2 \frac{\partial C_6}{\partial q}-\imath 2 \frac{\partial C_5}{\partial q}  \right) d
\label{eq:FJKUZB}
\mbox{,} \\
e&=&\left(H_1+\imath H_2 \right)d
\label{eq:FJKUZE}
\mbox{,} 
\end{eqnarray} 
where $H_1$ is an arbitrary real function of $q,\,t$, and  
$H_2$ is given by 
\begin{equation}
  \label{eq:AEKLF7}
H_2 =
2 m \tilde{u} 
\left( \frac{1}{\tilde{u}}\frac{\partial\tilde{u}}{\partial t}+\frac{\partial C_6}{\partial t} \right)
- 2 \frac{\partial C_5}{\partial q} \frac{\partial C_6}{\partial q}
- \frac{\partial^{2}C_5}{\partial q^{2}} 
\mbox{.}
\end{equation}
The quantity $\tilde{u}$, defined in Eq.~(\ref{eq:ABKUSCHL}), is an 
arbitrary function of $t$. The above relations show that all coefficients 
are proportional to $d$. Therefore, since $C_3=C_4=0$, the coefficient 
$d$ drops out of the differential equation. Thus, we are left with 
three arbitrary real functions, $C_5,\,C_6,\,H_1$, of $q,\,t$ and and a 
single arbitrary function $\tilde{u}$ of $t$. All other parameters 
may be expressed in terms of these four functions. 

The complex quantities $\chi$ and $F$ are given by
\begin{eqnarray}
  \chi &=& -\frac{\sqrt{\rho}}{\tilde{u}}\mathrm{e}^{-C_6}
\mathrm{e}^{\imath \left(\tilde{u} S+C_5 \right)}
   \label{eq:KMPCHIAN}
\mbox{,} \\
 F &=& \imath \frac{\sqrt{\rho}}{md}\mathrm{e}^{C_6}
\mathrm{e}^{-\imath \left(\tilde{u} S+C_5 \right)}
   \label{eq:DASFCAN}
\mbox{.}
\end{eqnarray}
Inserting the results for $a,\,b,\,e$ and dropping $d$
the differential equation for $\chi$ takes the form
\begin{equation}
  \label{eq:DGLICFAN}
\begin{split}
&\imath 2 m \tilde{u} \frac{\partial\chi}{\partial t}+
\left(2 \frac{\partial C_6}{\partial q}-\imath 2 \frac{\partial C_5}{\partial q} \right)
\frac{\partial\chi}{\partial q} + 
\frac{\partial^{2}\chi}{\partial q^{2}} \\
&+\bigg( H_1+
\imath \bigg[ 
2 m \tilde{u}
\left( \frac{1}{\tilde{u}}\frac{\partial\tilde{u}}{\partial t}+\frac{\partial C_6}{\partial t} \right)
- 2 \frac{\partial C_5}{\partial q} \frac{\partial C_6}{\partial q} - \frac{\partial^{2}C_5}{\partial q^{2}}
\bigg] 
\bigg)\chi = 0
\end{split}
\mbox{.}
\end{equation}   
From now on we use for simplicity the symbol for the partial derivative 
for all kinds of derivatives. At this point one may use the 
results~(\ref{eq:KMPCHIAN}),~(\ref{eq:DASFCAN}), to show that both 
sides of the fundamental condition~(\ref{eq:FUNDEQIA}) agree.

In order to compare with the previous, ``interaction-free'' 
equation~(\ref{eq:SCHRILY}) it is convenient to replace the 
fields $\tilde{u}$, $H_1$ by 
\begin{equation}
   \label{eq:KMJOUAN}
p(t)=\frac{1}{\tilde{u}}
\mbox{,}\;\;\;\;\;\; 
\tilde{V}=-\frac{H_1}{2 m \tilde{u}^{2}}
\mbox{.}
\end{equation}   
Using these fields and rearranging terms, Eq.~(\ref{eq:DGLICFAN}) 
takes the form
\begin{equation}
  \label{eq:DGLICNUF}
\begin{split}
& - \frac{p}{\imath} \frac{\partial\chi}{\partial t}+
2 \frac{p^{2}}{2 m }\left(\frac{\partial C_6}{\partial q}-\imath \frac{\partial C_5}{\partial q} \right)
\frac{\partial\chi}{\partial q} + 
\frac{p^{2}}{2 m } \frac{\partial^{2}\chi}{\partial q^{2}} \\
& -\tilde{V}\chi - \frac{p}{\imath}
\left(-\frac{1}{p} \frac{\partial p}{\partial t} + \frac{\partial C_6}{\partial t} \right)\chi
-\imath \frac{p^{2}}{2 m }
\left(2 \frac{\partial C_5}{\partial q} \frac{\partial C_6}{\partial q} + \frac{\partial^{2}C_5}{\partial q^{2}} \right)\chi 
= 0
\end{split}
\mbox{.}
\end{equation}
Comparing~(\ref{eq:DGLICNUF}) and (\ref{eq:SCHRILY})
one sees that now, instead of a single arbitrary function $V(q,t)$ 
in (\ref{eq:SCHRILY}), four arbitrary functions 
$p(t),\,\tilde{V}(q,t),\,C_6(q,t),\,C_5(q,t)$ appear 
in~(\ref{eq:DGLICNUF}). The field $p(t)$ has the dimension of 
an action and replaces $\hbar$, the field $\tilde{V}(q,t)$ replaces 
the previous mechanical potential $V(q,t)$. Each one of these four 
functions may play, in principle, the role of an ``agent'' for 
an external interaction. However it is possible (and we will see in 
a moment that this possibility applies) that some of these 
functions may be eliminated by means of a redefinition of the 
variable $\chi$. Let us try first to eliminate the functions $C_5,\,C_6$. 
Using the transformation 
\begin{equation}
  \label{eq:EDCCME}
\chi\Rightarrow\bar{\chi}=\chi\mathrm{e}^{C_6-\imath C_5}
\mbox{,}
\end{equation}
Eq.~(\ref{eq:DGLICNUF}) takes the form
\begin{equation}
  \label{eq:DGLINVEF}
- \frac{p}{\imath} \frac{\partial \bar{\chi}}{\partial t}+
\frac{p^{2}}{2 m } \frac{\partial^{2}\bar{\chi}}{\partial q^{2}} -V\bar{\chi} 
+ \frac{1}{\imath} \frac{\partial p}{\partial t}\bar{\chi} = 0
\mbox{,}
\end{equation}
if $\tilde{V}$ is replaced by $V$ according to the relation  
\begin{equation}
  \label{eq:AHFDVSV}
\tilde{V}=V-\frac{p^{2}}{2 m }
\bigg[
\left( \frac{\partial C_6}{\partial q}\right)^{2}-\left(\frac{\partial C_5}{\partial q} \right)^{2} 
+\frac{\partial^{2}C_6}{\partial q^{2}}
\bigg]
-p\frac{\partial C_5}{\partial t}
\mbox{.}
\end{equation}
Clearly, Eq.~(\ref{eq:AHFDVSV}) is permitted because both $\tilde{V}$ 
and $V$ are arbitrary functions. Eq.~(\ref{eq:DGLINVEF}) is a generalized 
Schr\"odinger equation, which agrees with the standard form for 
time-independent $p=\hbar$. Obviously, the functions $C_5,\,C_6$ are - in 
the framework of the present Ansatz - physically meaningless, because they 
may be eliminated from the dynamic equation by means of a simple 
redefinition of the state variable. 

If we use the arbitrary function $V$ instead of $\tilde{V}$, the 
untransformed equation~(\ref{eq:DGLICNUF}) takes the more familiar form
\begin{equation}
  \label{eq:DGLNAMAE}
 - \frac{p}{\imath} 
\big( \frac{\partial}{\partial t}-\frac{1}{p}\frac{\partial p}{\partial t}+\frac{\partial C_6}{\partial t}
-\imath \frac{\partial C_5}{\partial t} \big)\chi 
+ \frac{p^{2}}{2 m }
\big(\frac{\partial}{\partial q} + \frac{\partial C_6}{\partial q}
-\imath \frac{\partial C_5}{\partial q}\big)^{2}\chi
-V\chi=0 
\mbox{.}
\end{equation}
This equation is very similar to the one obtained from the
standard Schr\"odinger equation with the help of a gauge transformation;
the difference is (besides the time-dependent field $p$) that a  
complex function $C_5+\imath C_6$ appears instead of a real 
function $C_5$. In the present context, the gauge invariance of 
Schr\"odinger's equation is a consequence of the insensitivity of the
continuity equation against external disturbances. That part of the 
additional external $q,t$-dependence in $\chi$, that may be expressed by 
variable $C_5,\,C_6$ has been eliminated by the requirement that the 
continuity equation remains valid. As a consequence of this requirement
new ``compensating'' terms appeared in the coefficients. The present 
derivation opens a slightly different view on gauge theory. This point 
of view seems to be new, although the central role of the continuity 
equation and its symmetries for gauge theory is of course a 
well-established fact.

Can we continue this way and eliminate, as we did with $C_5$ 
and $C_6$, the arbitrary function $p(t)$ too from 
Eq.~(\ref{eq:DGLINVEF}), i.e. replace it by a constant $\hbar$ ?
This is apparently not the case. We may eliminate immediately the 
imaginary potential term (the one proportional to the time-derivative 
of $p$) from Eq.~(\ref{eq:DGLINVEF}) by means of the transformation
\begin{equation}
  \label{eq:EDHHI8}
\bar{\chi} \Rightarrow \chi_0= \bar{\chi} \mathrm{e}^{-\mathrm{ln}p}=\frac{\bar{\chi}}{p}
\mbox{.}
\end{equation}
But the result obtained this way namely
\begin{equation}
  \label{eq:DGNEEEF}
- \frac{p}{\imath} \frac{\partial \chi_0}{\partial t}+
\frac{p^{2}}{2 m } \frac{\partial^{2}\chi_0}{\partial q^{2}} -V\chi_0 = 0
\mbox{}
\end{equation} 
does not agree with Schr\"odinger's equation because still each
$\hbar$ in~(\ref{eq:SCHRILY}) is replaced by an arbitrary function
$p(t)$ in Eq.~(\ref{eq:DGNEEEF}). Of course, both equations agree 
if one sets approximately $p(t)\approx constant=\hbar$. Experimentally, this 
approximation seems to be valid; at least in the mostly investigated 
range of not too large time-intervalls. 

A time-dependence of Planck's constant $\hbar$ has been the subject
both of theoretical speculations and observations over large time 
periods~\cite{smith_oeztas_paul:model}. It is not clear at present
wether or not such a time-dependence exists. If it should turn out
to be real, this would rise the fascinating question of its physical 
origin; obviously it does not fit into the existing scheme to describe 
interactions. 

Let us recapitulate what we have done, paying particular attention to
the question of a time-dependent $\hbar$. In the last section we 
chose the simplest possible Ansatz to construct Schr\"odinger's equation.
We found an infinite number of equations characterized by three arbitrary 
constants, $p,\,C_5,\,C_6$, and an arbitrary function $V(q,t)$. Let 
us recall the slightly rewritten result for $\chi$, 
\begin{equation}
  \label{eq:KMPCHINAN}
\chi = - p \sqrt{\rho}\mathrm{e}^{-C_6+\imath C_5}
\mathrm{e}^{\imath \frac{S}{p}}
\mbox{.}
\end{equation}
In the present section we tried to make room for interactions without 
changing the basic framework. Essentially the only 
possibility to do that (see, however, the next section) was to 
allow for an arbitrary space-time dependence of the 
constants $p,\,C_5,\,C_6$. This is analogous to the usual 
procedure of ``gauging'', which means postulating a space-time dependence of 
the parameters of the gauge group. The standard gauging procedure 
leads then, in a next step, with the help of the concept of 
``compensating fields'', to the introduction of the gauge coupling 
terms. Clearly, the present approach may be considered as a generalized 
gauging procedure; the proceeding is basically the same for all 
three constants. One could argue - in favour of the reality of the 
time-dependent $\hbar$ -  that, in principle, this possibility to 
create interaction may be realized by nature for all parameters if 
it is realized for one of the constants (we know that it is realized 
for $C_6$, see the next section). On the other hand, the constants 
$C_5,\,C_6$ and $p$  in $\chi$ play a very different role [see 
Eq.~(\ref{eq:KMPCHINAN})]. Gauging of $C_5,\,C_6$ is compensated by 
additional terms in the coefficients, gauging of $p$ cannot be 
compensated completely. Only the first $p$ in Eq.~(\ref{eq:KMPCHINAN}), 
which plays, similar to $C_6$, the role of a scaling factor for the 
amplitude, may be eliminated. Combining the two above gauge 
transformations creates a state function, which is given by 
\begin{equation}
  \label{eq:ATRMHINAN}
\chi_0 =  \chi \mathrm{e}^{C_6-\imath C_5-\mathrm{ln}p} = - \sqrt{\rho}
\mathrm{e}^{\imath \frac{S}{p}}
\mbox{.}
\end{equation}
The variable $\chi_0$ still contains the field $p(t)$. The latter 
presents the only real ``interaction'' created by the extended 
Ansatz~(\ref{eq:ERWANS1}). It plays, as expected, the role of an  
arbitrary, time-dependent scaling factor for the action $S$.  
 
In the rest of this work we will neglect the time-dependence of $p$ 
and set $p=\hbar$. Then, in the framework of this approximation, 
the extended Ansatz~(\ref{eq:ERWANS1}) of the present section 
did \emph{not} lead to any real interactions and we conclude that 
at any rate a different, more general Ansatz is required to derive 
the form of the gauge coupling. This will be done in the next 
section, where all the results derived in the present section, 
will turn out to be useful. There, we will continue with the threedimensional 
generalization of~(\ref{eq:DGLNAMAE}), which [for $p(t)=\hbar$] is 
given by 
\begin{equation}
  \label{eq:DGMAENEQ}
 - \frac{\hbar}{\imath} 
\big( \frac{\partial}{\partial t}+\frac{\partial C_6}{\partial t}
-\imath \frac{\partial C_5}{\partial t} \big)\chi 
+ \frac{\hbar^{2}}{2 m }\sum_{i=1}^{3}
\big(\frac{\partial}{\partial q_i} + \frac{\partial C_6}{\partial q_i}
-\imath \frac{\partial C_5}{\partial q_i}\big)^{2}\chi
-V\chi=0 
\mbox{.}
\end{equation}
The state function $\chi$ depends on $q_i,\,t$ and is given by 
\begin{equation}
  \label{eq:DSEGMB2}
  \chi = -\hbar \sqrt{\rho}\mathrm{e}^{-C_6}
\mathrm{e}^{\imath \left(\frac{S}{\hbar}+C_5 \right)}
\mbox{.}
\end{equation}
We have (for $p=\hbar$) two possibilities to derive these equations. 
First, we expect that~(\ref{eq:DGMAENEQ}) and~(\ref{eq:DSEGMB2})
may be derived in a straightforward manner from a three-dimensional 
generalization of the fundamental condition~(\ref{eq:FUNDEQIA}). 
As a second way to derive Eq.~(\ref{eq:DGMAENEQ}) one may of course
perform simply a gauge transformation of the free Schr\"odinger equation. We 
will in the rest of this work  always refer to~(\ref{eq:DGMAENEQ}) 
and~(\ref{eq:DSEGMB2}) as derived according to the first possibility, 
i.e. from \emph{outside} quantum mechanics. The difference is important.
Choosing the first possibility, the role of the constants $C_5,\,C_6$ 
as ``agents'' of an external influence on the considered system is clear.
Choosing the second possibility, these constants are primarily abstract
group parameters without any direct physical significance (their role may, 
however, be suspected; e.g. one may show that $C_5$ does not depend on 
the wave function~\cite{kaempfer:concepts}).

\section{Second attempt to derive interaction}
\label{sec:5}
The plausible looking Ansatz of the last section was not successful. 
Trying to find a better one, it may be helpful to review first briefly 
the standard methods of introducing the (abelian) gauge field in quantum 
mechanics. Formally, the gauge field is introduced with the help of 
the minimal coupling rule
\begin{eqnarray}
  \frac{\partial}{\partial\vec{q}} &\Rightarrow& \frac{\partial}{\partial\vec{q}}-
\imath \frac{e}{\hbar c} \vec{A}
\label{eq:DWGK1}
\mbox{,} \\
  \frac{\partial}{\partial t} &\Rightarrow& \frac{\partial}{\partial t}+
\imath \frac{e}{\hbar} \phi
   \label{eq:DWGK2}
\mbox{,}
\end{eqnarray}
where $\vec{A}$ denotes the vector potential and $\phi$ the scalar 
potential. Applying~(\ref{eq:DWGK1}) and~(\ref{eq:DWGK2}) to 
Eq.~(\ref{eq:TDSCHR}) one obtains Schr\"odinger's equation for a 
particle in an electromagnetic field, 
\begin{equation}
  \label{eq:TDSCHREF}
-\frac{\hbar}{\imath}\frac{\partial}{\partial t}\psi-e\phi\psi=
-\frac{\hbar^{2}}{2m}
\left( 
\frac{\partial}{\partial\vec{q}}-\imath \frac{e}{\hbar c}\vec{A}
\right)^{2}\psi+V\psi
\mbox{.}
\end{equation}
If $\psi$ is again written in the form~(\ref{eq:PRODANS}) 
we obtain from~(\ref{eq:TDSCHREF}) the two equations
\begin{eqnarray}
\frac{\partial \rho}{\partial t}&+&\frac{\partial}{\partial\vec{q}}
\frac{\rho}{m} \left( 
\frac{\partial S}{\partial\vec{q}}-\frac{e}{c}\vec{A}
\right)=0
  \label{eq:CONTMF}
\mbox{,} \\
\frac{\partial S}{\partial t}+e\phi +V &=&-\frac{1}{2m}
\left(
\frac{\partial S}{\partial\vec{q}}-\frac{e}{c}\vec{A}
\right)^{2}+
\frac{\hbar^2}{2m}\frac{\triangle\sqrt{\rho}}{\sqrt{\rho}}
\label{eq:QHJMF}
\mbox{.} 
\end{eqnarray}  
Using the variables $\rho$ and $S$, there is also a minimal coupling 
rule to create the interaction terms in~(\ref{eq:CONTMF}) 
and~(\ref{eq:QHJMF}) from the interaction-free 
equations~(\ref{eq:CONT}) and~(\ref{eq:QHJ}). It is 
obviously given by
\begin{eqnarray}
  \frac{\partial S}{\partial\vec{q}} &\Rightarrow& \frac{\partial S}{\partial\vec{q}}-\frac{e}{c} \vec{A}
\label{eq:DWGKRS1}
\mbox{,} \\
  \frac{\partial S}{\partial t} &\Rightarrow& \frac{\partial S}{\partial t}+ e \phi
   \label{eq:DWGKRS2}
\mbox{.}
\end{eqnarray} 
How to justify the minimal coupling rule ? The standard explanation 
is the ``principle of local gauge invariance''. It requires that 
the parameters of the global symmetry group (the constants $C_5,\,C_6$
in our notation) be arbitrary functions of $q_k,\,t$ without destroying 
the form invariance of Schr\"odinger's equation. This requires, thus the 
line of argument, introduction of a compensating field, the gauge field. 
However, as Eq.~(\ref{eq:DGLNAMAE}) shows, Schr\"odinger's equation \emph{is} 
already invariant under the local group; no matter wether or not the new 
fields have any physical effect. This has been pointed 
out~\cite{ogievetski:gauge} a few years after publication of the 
fundamental papers~\cite{yang-mills:gauge}, \cite{utiyama:invariant}
on gauge theory. The principle of gauge invariance alone, without 
additional assumptions, cannot explain the minimal 
coupling rule.

A second method to introduce a gauge field exists. There one postulates 
that the phase of the wave function (or rather a part of it) is 
non-integrable. This method goes back to Weyl~\cite{weyl:elektron} and 
London~\cite{london:gauge} and is discussed in many works. We mention only 
two, a compact presentation to be found in Dirac's monopol 
paper~\cite{dirac:quantised} and a more detailed and clearly written 
discussion in a book by Kaempfer~\cite{kaempfer:concepts}. The basic 
idea of the non-integrable phase may be explained very quickly by 
starting from Eq.~(\ref{eq:TDSCHREF}) and eliminating the potentials by means 
of a singular gauge transformation~\cite{kaempfer:concepts}. The result is 
an equation
\begin{equation}
  \label{eq:SCHRILYNP}
-\frac{\hbar}{\imath}\frac{\partial\psi_{\mathcal{C}}}{\partial t}=
-\frac{\hbar^{2}}{2m}\frac{\partial^{2}\psi_{\mathcal{C}}}{\partial q^{2}}+
V(q,t)\psi_{\mathcal{C}}
\mbox{,}
\end{equation}
which looks like the free Schr\"odinger equation but cannot be used as
a differential equation because the wave function $\psi_{\mathcal{C}}$ 
is a multi-valued mathematical object with non-integrable phase; 
the notation $\psi_{\mathcal{C}}$ indicates an unspecified dependence 
on a path $\mathcal{C}$. The gauge field may be introduced exactly 
the other way round~\cite{dirac:quantised}. One starts from the free 
Schr\"odinger equation, postulates the existence of a non-integrable 
phase and eliminates this singular part by means of a gauge transformation, 
thereby creating the potentials in Eq.~(\ref{eq:TDSCHREF}). 
How to justify the introduction of the non-integrable phase ? 
Apparently, this question has not been discussed in the literature. 
This is surprising since the fact that Eq.~(\ref{eq:SCHRILYNP}) takes 
the form of the free Schr\"odinger equation is very remarkable and calls 
for an explanation. 

This second method to introduce a gauge field, by means of a non-integrable 
phase, may be adapted for the present problem. A possible explanation - which 
is however still speculative in character - for the non-uniqueness of the 
phase will also be given (in the next section). Adapting this idea means 
that we allow in the Ansatz~(\ref{eq:ERWANS1}) for a \emph{non-unique} 
explicit space-time dependence of the state function $\chi$. Introducing a 
non-unique mathematical object in a physical theory implies more or less 
automatically a further requirement, namely that all mathematical quantities 
of direct physical significance are well-behaved (unique) functions. Such a 
requirement (which is for well-behaved functions obvious and not worth 
mentioning), should be formulated  explicitly in the present situation. 
We require, in particular, that the state function $\chi$ itself is a unique 
(up to regular gauge transformations) function of $q,\,t$ and that all 
derivatives are unique functions of $q,\,t$; these requirements imply  
that we obtain a well-defined differential equation with unique 
coefficients. Thus, our once more extended Ansatz takes the form 
\begin{eqnarray}
  &\chi&=\chi(\rho,S;q,t)
   \label{eq:DNEADE1}
\mbox{,} \\
&\chi&\; \mbox{unique},\;\;\mbox{all derivatives of}\; \chi\; \mbox{unique}
   \label{eq:DNEADZ2}
\mbox{,}
\end{eqnarray}
where the semicolon in~(\ref{eq:DNEADE1}) has been introduced to 
indicate the allowed non-uniqueness of the explicit $q,\,t$-dependence. 
It should be pointed out that the non-unqueness is allowed but 
not required. It will be realized only if the consistency 
requirements~(\ref{eq:DNEADZ2}) can be fullfilled (how this 
actually works will become clear in a moment).

We have to construct, using~(\ref{eq:DNEADE1}),~(\ref{eq:DNEADZ2}), 
a state equation for $\chi$. The basic construction scheme 
is the same as in the last section; in particular we start,
from a fundamental condition which agrees formally with 
~(\ref{eq:FUNDEQIA}). Thus we have to solve 
\begin{equation}
  \label{eq:FUNEQIBB} 
\Re\left[F\left(a\frac{\mathrm{d}}{\mathrm{d}t}+
b\frac{\mathrm{d}}{\mathrm{d} q}+
\mathrm{d}\frac{\mathrm{d}^{2}}{\mathrm{d}q^{2}}+
e\right)\chi\right]=\frac{\partial \rho}{\partial t}+
\frac{\partial \rho}{\partial q}\frac{1}{m}\frac{\partial S}{\partial q}+
\rho\frac{1}{m}\frac{\partial^{2}S}{\partial q^{2}}
\mbox{,} 
\end{equation}
together with the conditions~(\ref{eq:DNEADE1}) 
and~(\ref{eq:DNEADZ2}). The problem defined by~(\ref{eq:DNEADE1}) - 
(\ref{eq:FUNEQIBB}) is in principle completely independent from the 
problems dealt with in the last two sections. But all the formal mathematical
results, obtained in the last two sections, may be taken over to the 
present problem; the new aspect - the non-uniqueness - does not change 
these results. The non-unique $q,\,t$-dependence of $\chi$ may be expressed 
by three non-unique functions, namely $p(t)$, $C_5(q,t)$, $C_6(q,t)$, and the 
derivatives of these functions have to be, according to 
condition~(\ref{eq:DNEADZ2}) unique functions of $q,\,t$. Thus we 
may immediately conclude that Eq.~(\ref{eq:FUNEQIBB}) implies 
relation~(\ref{eq:DGLNAMAE}). The latter will become a well-defined 
differential equation, if we are able to transform the non-unique 
quantity appearing in this equation [also called $\chi$ in~(\ref{eq:DGLNAMAE})] 
by proper manipulations into a unique function of $q,\,t$. In other 
words, the only remaining problem is to fulfill condition~(\ref{eq:DNEADZ2}). 
Before we proceed we neglect, as discussed in the last section, 
in~(\ref{eq:DGLNAMAE}) the time-dependence of $p(t)$ and set $p(t)=\hbar$. 
Further, we shall use in the following the threedimensional Schr\"odinger 
equation~(\ref{eq:DGMAENEQ}) instead of ~(\ref{eq:DGLNAMAE}). This is 
because the structures to be studied can only be represented adequately 
in a fourdimensional space-time continuum. 

Our starting point is Eq.~(\ref{eq:DSEGMB2}); obviously, 
condition~(\ref{eq:DNEADZ2}) is not yet fulfilled. It is also clear 
that the functions $C_5$ and $C_6$ may be treated separately; 
given that we are unable to implement uniqueness, we may as well 
eliminate one or both of these functions by means of a singular gauge
transformation (or simply set it equal to zero). Now all relations 
between our functions are fixed already; there is only a single 
possibility left to implement an additional condition: The 
functions $\rho$ and $S$ themselfes may be defined to be non-unique
functions of $q_i,\,t$.

First we assume that $C_6$ cannot be implemented as a non-unique
function and set $C_6=0$~ in~(\ref{eq:DGMAENEQ}) and (\ref{eq:DSEGMB2}).
The form of $\chi$ shows, that the remaining non-uniqueness of $C_5$ may 
be compensated by postulating a non-unique variable $S$ according to 
\begin{equation}
  \label{eq:DFEPD1}
S=-\hbar C_5 + \hbar \varphi
\mbox{,}
\end{equation}
where $\varphi$ is a unique (up to regular gauge transformations) function of 
$q_i,\,t$. This simple linear combination eliminates the non-uniqueness 
of $C_5$ by means of a non-unique $S$ in favor of a unique $\varphi$ and 
produces a unique state function $\chi$. If, instead of $\varphi$, a quantity 
with the dimension of an action is required, we write $\varphi=\bar{S} /  \hbar$. 
Eq.~(\ref{eq:CONT}) shows that the replacement of  $S$ by $\bar{S}$ 
according to 
\begin{equation}
  \label{eq:ENPMUVF}
S \longrightarrow \bar{S}=S + \hbar C_5
\mbox{}
\end{equation}
is compatible with the structure of the continuity equation and 
leads only to a redefinition of the probability current. Otherwise, 
this compensation procedure would not make sense.

Let us first follow the consequences of a non-unique $C_5$ before 
we come back to $C_6$. In order to obtain an explicit representation 
for $C_5$ we introduce for its derivatives, which are unique functions 
according to~(\ref{eq:DNEADZ2}), the following symbols,
\begin{equation}
  \label{eq:DAVSPSE}
\frac{\partial C_5}{\partial q_k}=\bar{A}_k,\;\;\;\frac{\partial C_5}{\partial t}=\bar{\phi}
\mbox{.}
\end{equation}
Then, $C_5$ may be written as an integral over a path $\mathcal{C}$, 
\begin{equation}
  \label{eq:AWIMMVF}
C_5(\vec{q},t;\mathcal{C})=
\int_{\vec{q}_{0},t_{0};\mathcal{C}}^{\vec{q},t}
\left[\mathrm{d}q_k' \bar{A}_k(\vec{q}',t')+ \mathrm{d}t' \bar{\phi}(\vec{q}',t')\right]
\mbox{.}
\end{equation}
The last formula (summation over double indices) represents one of the 
fundamental postulates of gauge theory~\cite{weyl:elektron},~\cite{dirac:quantised}, ~\cite{kaempfer:concepts}. The multi-valuedness of $C_5$ is expressed 
by the fact that $C_5$ does not only depend on the considered space-time 
point $\vec{q},t$ but also on the path $\mathcal{C}$ which leads from a 
(fixed) reference point $\vec{q}_{0},t_{0}$ to $\vec{q},t$.
All conceivable values of $C_5$ may be obtained by specifying 
four real fields 
$\bar{\phi}(\vec{q},t)$, $\bar{A}_k(\vec{q},t),\,k=1,2,3$. 

These four fields are not independent from each other. In the 
following we reproduce some well-known 
results~\cite{dirac:quantised},\cite{kaempfer:concepts} which are 
essential for the physical interpretation of these fields. We 
define a four-vector $x_{\mu},\;\mu=0,1,2,3$ by  $x_{0}=v_0 t$, (where 
$v_0$ is an unknown constant with dimension of a velocity) and 
$x_{k}=q_k,\;k=1,2,3$. The four-vector field $\tilde{A}_{\mu}$ is 
defined by $\tilde{A}_{0}(x_{\mu})=\bar{\phi}(q_k,t)/v_0$ and  
$\tilde{A}_{k}(x_{\mu})=\bar{A}_k(q_k,t),\;k=1,2,3$.
Then, $C_5$ may be written in the form
\begin{equation}
  \label{eq:AWIMSRF}
C_5(\vec{q},t;\mathcal{C})=\tilde{C}_5(x;\mathcal{C})=
\int_{x_{\mu,0};\mathcal{C}}^{x_{\mu}}
\mathrm{d}x_{\mu}' \tilde{A}_{\mu}(x')
\mbox{.}
\end{equation}
Integrating along an arbitrary closed path $\mathcal{C}_0$ 
from $x_{\mu}$ to $x_{\mu}$ Stokes 
integral theorem implies 
\begin{equation}
  \label{eq:D4DIMST}
\oint_{\mathcal{C}_0} \mathrm{d}x_{\mu}' \tilde{A}_{\mu}(x')=
\int_{\mathcal{A}(\mathcal{C}_0)} \mathrm{d}f_{\mu\nu}'
\left(\frac{\partial\tilde{A}_{\nu}}{\partial x_{\mu}'}-\frac{\partial\tilde{A}_{\mu}}{\partial x_{\nu}'} \right)
\mbox{,}
\end{equation}
where the antisymmetric tensor $\mathrm{d}f_{\mu\nu}$ characterizes
the infinitesimal surface element of a surface $\mathcal{A}(\mathcal{C}_0)$, 
which is bounded by our curve $\mathcal{C}_0$ and otherwise arbitrary. 
Given that $C_5$ is multi-valued, the path integral on the left side of 
Eq.~(\ref{eq:D4DIMST}) must not vanish for arbitrary paths 
$\mathcal{C}_0$. This implies that the integrand of the surface 
integral, which is denoted by $\tilde{F}_{\mu\nu}$, must not vanish for 
arbitrary space-time points. In other words, the set of points 
defined by 
\begin{equation}
  \label{eq:DLAFSTTA}
\tilde{F}_{\mu\nu}\equiv\frac{\partial\tilde{A}_{\nu}}{\partial x_{\mu}}-\frac{\partial\tilde{A}_{\mu}}{\partial x_{\nu}}\neq0
\mbox{,}
\end{equation}
must not be empty. At the points where Eq.~(\ref{eq:DLAFSTTA}) holds, 
the non-uniqueness of $C_5$ implies the non-commutativity of the 
derivatives with regard to $x_{\mu}$. If $\tilde{F}_{\mu\nu}=0$ for all $x_{\mu}$, 
then the influence of $\tilde{A}_{\mu}$ (and $C_5$) may be eliminated 
by means of a regular gauge transformation. If, on the other hand, 
points exist where Eq.~(\ref{eq:DLAFSTTA}) holds true, then the values 
$\tilde{F}_{\mu\nu}$ takes at these points are invariant under regular 
gauge transformations. This means that $\tilde{F}_{\mu\nu}$ has a gauge 
invariant meaning and implies the possibility that $\tilde{F}_{\mu\nu}$ 
plays a role as (is proportional to) a classical field or force.

As a consequence of the definition~(\ref{eq:DLAFSTTA})
the field $\tilde{F}_{\mu\nu}$ obeys the differential equation
\begin{equation}
  \label{eq:DEFDHMG}
T_{\lambda\mu\nu}\equiv\frac{\partial\tilde{F}_{\mu\nu}}{\partial x_{\lambda}}+
\frac{\partial\tilde{F}_{\nu\lambda}}{\partial x_{\mu}}+
\frac{\partial\tilde{F}_{\lambda\mu}}{\partial x_{\nu}}=0
\mbox{.}
\end{equation}
The left hand side $T_{\lambda\mu\nu}$ of Eq.~(\ref{eq:DEFDHMG}) is antisymmetric 
in all three indices. Only four relations in Eq.~(\ref{eq:DEFDHMG})
are independent. These may be written in the form
\begin{equation}
  \label{eq:AKFFDLHG}
\varepsilon_{\kappa\lambda\mu\nu}\frac{\partial\tilde{F}_{\mu\nu}}{\partial x_{\lambda}}=0
\mbox{.}
\end{equation}
The definition of $\varepsilon_{\kappa\lambda\mu\nu}$  may be found e.g. in the 
book by Landau~\cite{landau.lifshitz:classical}. 
If the essential components
of $\tilde{F}_{\mu\nu}$ are renamed according to
\begin{equation}
  \label{eq:DBFDKDFT}
\begin{split}
&\tilde{F}_{01}=-\tilde{E}_{1},\;\;\;
\tilde{F}_{02}=-\tilde{E}_{2},\;\;\;
\tilde{F}_{03}=-\tilde{E}_{3},\;\;\; \\
&\tilde{F}_{12}=+\tilde{B}_{3},\;\;\;
\tilde{F}_{13}=-\tilde{B}_{2},\;\;\;
\tilde{F}_{23}=+\tilde{B}_{1}
\end{split}
\mbox{,}
\end{equation}
then the four relations~(\ref{eq:AKFFDLHG}) may be written 
as differential equations for the vektorfields 
$\vec{\tilde{E}}$ and $\vec{\tilde{B}}$ (with components 
$\tilde{E}_{i}$ and $\tilde{B}_{i}$), 
\begin{equation}
  \label{eq:DMGINVF1}
\frac{\partial\vec{\tilde{B}}}{\partial\vec{r}}=0,\;\;\;\;\;\;
\frac{\partial}{\partial\vec{r}} \times \vec{\tilde{E}} 
+\frac{1}{v_0} \frac{\partial\vec{\tilde{B}}}{\partial t}=0
\mbox{.}
\end{equation}
Thus, the fields $\tilde{E}_{i},\;\tilde{B}_{i}$ must be solutions of 
the homogeneous Maxwell equations, if the constant $v_0$ is 
identified with the velocity of light $c$. Of course, 
Eqs.~(\ref{eq:DMGINVF1}) are not sufficient to determine 
$\tilde{E}_{i},\;\tilde{B}_{i}$; some more equations are required. 
An obvious possibility is to identify $v_0$ with $c$ and to postulate
that the rest of the equations is given by the two inhomogeneous 
Maxwell equations, i.e. to assume that $\vec{\tilde{E}}$ and 
$\vec{\tilde{B}}$ are proportional to the electric and magnetic 
field vectors $\vec{E}$ und $\vec{B}$ of Maxwell's theory.
This postulate, which is an essential part of Hermann Weyls 
first gauge theory, is closely related  to the structure of 
Eq.~(\ref{eq:AWIMMVF}). 

We still have to find constants of proportionality with suitable 
dimensions. The ``fields'' $\tilde{E}_{i}$ and $\tilde{B}_{i}$ (both 
with dimension $cm^{-2}$) have to be connected, by means of proper 
constants of proportionality, to the two basic terms of the classical 
particle-field concept, \emph{force} and \emph{field}. If a field 
$\bar{E}_{i}$, which has the dimension of a force, is defined by 
means of $\alpha\tilde{E}_{i}=\bar{E}_{i}$, then the constant $\alpha$ has 
the dimension $g^{1} cm^{3} sec^{-2}$. Using the available 
constants $m$, $\hbar$, $c$ only a single combination with suitable 
dimension may be formed, namely $\hbar c$. So we set $\alpha=\hbar c$. 
In order to have an ``objective'' field $E_{i}$, whose existence 
does not depend on the presence or absence of a test particle, we have 
to introduce one more constant, the charge $e$. It is, like the mass $m$, 
part of the description of the individual particle. Thus we write 
$\bar{E}_{i}=e E_i$, where $e$ has the dimension 
$g^{\frac{1}{2}} cm^{\frac{3}{2}} sec^{-1}$. Combining both 
constants, and making similar considerations for the 
magnetic field, we obtain
\begin{equation}
  \label{eq:DZZDGUDA}
\tilde{E}_{i}=\frac{e}{\hbar c} E_{i},\;\;\;\;\;\;
\tilde{B}_{i}=\frac{e}{\hbar c} B_{i},\;\;\;\;\;\;
\tilde{F}_{\mu\nu}=\frac{e}{\hbar c} F_{\mu\nu}
\mbox{.}
\end{equation}
Similar constants of proportionality occur if the potentials 
$\tilde{A}_{\mu}$ are replaced  by the standard 
potentials of electrodynamics $A_k,\,\phi$, 
\begin{equation}
  \label{eq:ADPMFM2}
\tilde{A}_{0}(x_{\mu})=-\frac{e}{\hbar c} \phi(q_k,t),\;\;\;\;\;\;
\tilde{A}_{k}(x_{\mu})=\frac{e}{\hbar c} A_k(q_k,t)
\mbox{.}
\end{equation}
Replacing in an analogous way $C_5$ by $C$
with the help of the relation 
\begin{equation}
  \label{eq:DFA3UER}
C_5(\vec{q},t;\mathcal{C})=\frac{e}{\hbar c}
C(\vec{q},t;\mathcal{C})
\mbox{,}
\end{equation}
the multi-valued function $C$ may be written as a path integral with 
the usual potentials appearing in the integrand 
\begin{equation}
  \label{eq:ALLO4VF}
C(\vec{q},t;\mathcal{C})=
\int_{\vec{q}_{0},t_{0};\mathcal{C}}^{\vec{q},t}
\left[\mathrm{d}q_k' A_k(\vec{q}',t')- c \mathrm{d}t' \phi(\vec{q}',t')\right]
\mbox{.}
\end{equation}
Using these properly scaled variables, we obtain the following well-known
relation between the components of $F_{\mu\nu}$ and the potentials $A_k,\,\phi$,
\begin{equation}
  \label{eq:AGEMWJ7}
\vec{E}=-\frac{1}{c}\frac{\partial\vec{A}}{\partial t}-\frac{\partial\phi}{\partial\vec{r}},\;\;\;\;\;\;
\vec{B}=\frac{\partial}{\partial\vec{r}} \times \vec{A}
\mbox{.}
\end{equation}
Finally, with the help of the relations 
\begin{equation}
  \label{eq:GERWE8U}
\frac{\partial C_5}{\partial q_k}=\frac{e}{\hbar c} A_k,\;\;\;\;\;\;
\frac{\partial C_5}{\partial t}=-\frac{e}{\hbar } \phi
\mbox{,}
\end{equation}
(and with $C_6=0$) Schr\"odinger's equation~(\ref{eq:DGMAENE}) 
for the variable
\begin{equation}
  \label{eq:WFSCHRNBSA}
\psi  =-\chi / \hbar = \sqrt{\rho}\mathrm{e}^{\imath \frac{\bar{S}}{\hbar}},
\;\;\;\;\;\;\;\;
\bar{S}=S+\frac{e}{c}C
\mbox{,}
\end{equation}
takes the form
\begin{equation}
  \label{eq:DGMAENE}
 - \frac{\hbar}{\imath} 
\big( \frac{\partial}{\partial t}+\imath \frac{e}{\hbar } \phi \big)\psi 
+ \frac{\hbar^{2}}{2 m }\sum_{i=1}^{3}
\big(\frac{\partial}{\partial q_i} -\imath \frac{e}{\hbar c} A_k \big)^{2}\psi
-V\psi=0 
\mbox{.}
\end{equation}
Thus, our second attempt to derive Schr\"odinger's equation with gauge 
coupling turned out to be successful. 

Now we come back to the question if a meaningful theory may be 
constructed with a nonzero multi-valued $C_6$. There is a 
standard argument on this point in the text-book literature. It says 
(using the present notation) that a nonzero $C_6$ is forbidden, 
because it leads (in contrast to a phase change) to a modification 
of the amplitude of the wave function and to a corresponding change 
in probability density. If this argument were true, a modification 
of the phase would be forbidden as well. The latter leads also to 
a modification of the amplitude - as a consequence of the coupling 
between phase and amplitude. However, while this argument is strictly 
speaking wrong it leads to the correct conclusion. We may proceed as 
we did with $C_5$, allowing for a multi-valued $\rho$ and defining a 
new probability density $\bar{\rho}$ by means of the relation  
\begin{equation}
\label{eq:NWDIF4}
\bar{\rho}=\rho \mathrm{e}^{-2C_6}
\mbox{.}
\end{equation}
The functions $\rho$ and $C_6$ are multi-valued while 
$\bar{\rho}$ is single-valued. We may represent $C_6$ again 
(as we did with $C_5$) as a path integral and the associated 
potentials would again fulfill the homogeneous Maxwell equations.
But this is not sufficient. In order for the substitution 
$\rho\to\bar{\rho}$ (together with $S\to\bar{S}$) to make sense, it is
necessary that the structure of the continuity equation 
remains intact. This means that performing the substitution 
$\rho\to\bar{\rho}$ in 
\begin{equation}
  \label{eq:CONTMSQ4}
\frac{\partial \rho}{\partial t}+\frac{\partial}{\partial\vec{q}}
\frac{\rho}{m} 
\left( \frac{\partial \bar{S}}{\partial\vec{q}}-\frac{e}{c}\vec{A} \right)
=0 
\mbox{}
\end{equation}
produces a mathematically well-defined differential equation 
with unique coefficients, which still has the structure of 
a continuity equation (with possibly redefined density and 
current). This is not the case as insertion of~(\ref{eq:NWDIF4}) 
in ~(\ref{eq:CONTMSQ4}) shows. Only the interaction mediated
by a non-unique $C_5$ can be realized in nature.  
  
\section{Discussion}
\label{sec:6}
In this section we will give a summary of the paper and discuss its 
most important results. The calculations performed in section~\ref{sec:3}) 
led us to conjecture that Schr\"odinger's equation may be derived from the 
following three assumptions 
\begin{itemize}
\item [(1)] The continuity equations holds for a 
probability density $\rho$ and a probability current $\vec{j}$, 
which depends linearly on the gradient of a function $S$. 
\item [(2)]
The  system may be described by a complex state function $\chi$.
\item [(3)]
The state function $\chi$ obeys a linear differential 
equation.
\end{itemize}
We have not verified in detail this conjecture but given arguments 
supporting its validity in section~\ref{sec:3}); in the following 
discussion it will be considered as true. 

The above three assumptions define a quantization procedure. The 
system that has been quantized was, however, not a particle but 
rather a statistical ensemble of particles. This quantization method 
does not require a Hamiltonian; instead the conservation law of 
probability (first assumption) represents the fundamental input 
of this theory, which may be roughly characterized as a statistical
quantization method. The linearity (third assumption) of the 
state equation may also be understood in terms of the statistical 
nature of this theory. Finally, the second assumption of a complex-valued 
state function is the essential ``non-classical'' part of this 
theory. This is a formal requirement and its physical meaning and 
origin is a \emph{priori} unclear. It is, however, a very simple assumption, 
much simpler than e.g. the canonical commutation relations, which are 
the essential nonclassical part of the conventional quantization method 
and are also of a formal nature. 

We may ask why the present quantization method is conceptually 
simpler than the conventional one. The present method starts from
a statistical ensemble while the conventional method starts from a 
single particle. Let us consider the classical limit of Schr\"odinger's
equation. This limit may be performed conveniently by setting $\hbar=0$ 
in Eqs.~(\ref{eq:CONT}) and~(\ref{eq:QHJ}). The result is the Hamilton-Jacobi 
equation~(\ref{eq:HJCL}) \emph{and} the continuity equation~(\ref{eq:CONT}). 
The latter is obviously not eliminated by performing the limit $\hbar\Rightarrow0$ 
because it does not contain the constant $\hbar$. The classical 
limit~(\ref{eq:HJCL}),~(\ref{eq:CONT}) of Schr\"odinger's equation is a 
field theory, which describes an infinite number of particle trajectories. 
Each one of these trajectories may be calculated for given initial 
conditions in the framework of classical point mechanics [there is no 
coupling to $\rho$ in Eq.~(\ref{eq:HJCL})] , but these trajectories are 
only realized with a certain probability which must be calculated with 
the help of~(\ref{eq:CONT}). The classical limit of Schr\"odinger's 
equation is not a particle theory but a statistical theory; this lends 
support to the ensemble interpretation of quantum 
theory~\cite{home:ensemble}. Therefore, for the inverse problem, the 
transition from classical physics to quantum theory, a statistical law 
may be a simpler (more natural) starting point than a particle law. 

Three different functional forms for the dependence of $\chi$ on $\rho$ 
and $S$ have been used in sections~\ref{sec:3}),~\ref{sec:4}) 
and~\ref{sec:5}). The first, simplest Ansatz $\chi=\chi(\rho,\,S)$, in 
section~\ref{sec:3}), has been called ``interaction free''.  
The result was Schr\"odinger's equation for a particle (ensemble) in
an external mechanical potential $V(q,t)$. 

In the second Ansatz $\chi=\chi(\rho,\,S,\,q,\,t)$, in section~\ref{sec:4}), 
the possibility of an explicit dependence of $\chi$ on $q,t$ was taken 
into account. Each constant derived in section~\ref{sec:3}) becomes a 
possible ``channel'' for this additional $q,\,t$-dependence. One of 
these channels leads to a new and rather exotic ``influence'' on our 
system, which is formally described by a time-dependent Planck 
constant $\hbar$. All other channels are eliminated by additional 
compensating terms in the coefficients.  By means of a simple redefinition 
of the state variable the resulting equation may (for neglegible 
variation of $\hbar$) be transformed to the free Schr\"odinger equation 
of section~\ref{sec:3}). This tranformation agrees exactly with the 
usual quantum mechanical gauge transformation. Therefore, the Ansatz 
$\chi=\chi(\rho,\,S,\,q,\,t)$ contains the usual gauging procedure as a special 
case. The gauge coupling terms itself cannot be derived within this 
Ansatz; it turns out, however, that it presents nevertheless an 
essential step towards its introduction.
 
The third Ansatz in section~\ref{sec:5}) has been written in 
the form $\chi=\chi(\rho,\,S;\,q,\,t)$, which means that a multi-valued external 
space-time dependence of $\chi$ is permitted. The idea for this Ansatz 
stems from the well-known quantum mechanical concept of a 
non-integrable phase. The functional form of all dependencies is 
already fixed by the calculations of section~\ref{sec:4}). It 
turns out that the only channel leading to an actual physical 
effect is given by the parameter $C_5$. Its multi-valuedness must 
be compensated for by the multi-valuedness of the quantity $S$ (the 
generalization of the classical action), in order for the final 
phase of $\chi$ to be well-behaved. Performing a variable substitution 
one obtains finally the correct gauge coupling terms in 
Schr\"odinger's equation. In the standard formulation of gauge theory,
the compensation effect is meant to restore the (locally destroyed) 
gauge symmetry, in the present formulation it restores the 
uniqueness of the state function. The last step to the introduction
of the gauge field, performed in section~\ref{sec:5}), is formally 
nearly identical to the standard theory of non-integrable phases. 
There is, however, an important conceptual difference. The 
field $C_5(q,t)$ of the present theory represents by definition an 
externally controlled influence on the considered system. The 
multi-valuedness of the action $S(q,t)$, which describes the system, 
is a consequence of the multi-valuedness of $C_5(q,t)$. The latter 
may be identified as reason while the former is the effect of the 
latter. This clear cause-reason relation is absent in the standard 
theory of the non-integrable phase. 

Let us come back to the question raised in section~\ref{sec:5}) how 
to justify, or ``understand'', the form of the gauge coupling. 
We know that the non-uniqueness of $C_5(q,t)$, which describes 
some external influence on our system, is a necessary condition 
for the occurrence of a gauge field. So, what is the reason for the 
non-uniqueness of $C_5(q,t)$; where does it come from ? The 
following is not a final answer to this question but rather a 
coherent collection of remarks intended to stimulate further research. 
 
The non-uniqueness of a physical quantity means that it is 
impossible to express its effect in a ``local'' way, by stating 
its value on a particular space-time point. The term ``local'' needs 
further explanation. It applies practically to all of the conventional 
scheme of formulating physical laws by means of differential equations; 
in this scheme all the information required to predict (or retrodict) 
the behavior of a system is determined by stating its values (e.g. 
coordinates of particles or field values at all space points) at a 
particular instant of time. There is no universal law dictating that 
this ``local'' conventional scheme must be always true. It may well be 
that instead the values of the variables in a time intervall, of 
finite extent, are required to predict the behavior of a system. 
Mathematically such theories are described e.g. by delay differential 
equations. 

Let us consider the classical theory of charged particles and fields,
having in mind a possible breakdown of the conventional scheme. 
Nearly all applications of this theory study one of two idealized
situations. The first is calculating the fields produced by given sources,
the second is calculating the trajectories of particles in given fields.
Both types of problems are, strictly speaking, unrealistic, even if
this does not cause any problems in macroscopic situations. Realistic 
problems have to take into accout the mutual influence of trajectories 
and fields. Whenever this is done 
seriously~\cite{raju:electrodynamic},~\cite{deluca:geometric},~\cite{chicone.ea} one encounters systems of delay-differential equations 
which are difficult to solve and whose mathematical structure is  
still to be analyzed in detail. But if one tries to simulate a 
realistic ``nonlocal'' problem in the framework of the conventional 
scheme, one encounters necessarily problems, in the form of 
non-unique or multi-valued predictions. This is what happens already 
for the simplest conceivable realistic problem of this type, a charged 
particle in an external electromagnetic field under the influence of 
its own radiation field~\cite{rohrlich:classical}, \cite{spohn:manifold_lorentz_dirac}. 

Now, quantum mechanics (as well as the present theory) uses, of course,
the conventional ``local'' scheme of formulating physical laws. It is possible
that the above explained ``nonlocality'' is - in a way still to be clarified 
in detail - responsible for the non-uniqueness of  $C_5(q,t)$. The latter 
plays, according to the derivation reported in section~\ref{sec:5}), a 
central role for the form of the gauge coupling. May we extend this idea 
about the origin of the form of the gauge coupling further, to quantum 
mechanics itself ? The above calculation was in fact a simultaneous 
derivation of the fundamental equation of quantum theory \emph{and} the 
form of the gauge coupling. Considering it that way, we made three (or 
two, if we neglect the possibility of a time-dependent Planck constant) 
different derivations of quantum theory, and may ask which one is most 
fundamental. The third derivation led to Schr\"odinger's equation with the 
minimal coupling rule, which is obeyed by all interactions found in nature 
(including non-abelian generalizations and possibly excluding gravity). It  
should be considered as most fundamental, if quantum mechanics is to be 
understood as a step towards a realistic description of nature. If one 
adopts this point of view, Schr\"odinger's equation depends also crucially 
on the assumption of a non-unique external influence. We conclude that a 
detailed study of the breakdown of the ``local'' formulation of realistic 
classical physics may be useful in order to achieve a deeper understanding 
of quantum mechanics.

This attitude towards quantum theory is not new, although a minority view.
We mention, in particular, the work of Raju~\cite{raju:electrodynamic}.
He revealed clearly crucial properties of classical particle-field systems 
which have been overlooked by the scientific community for nearly a century. 
Of course, such findings are also relevant for a careful 
characterization of quantum mechanical nonlocality. Nelson, using a 
quite different starting point, conjectured~\cite{nelson:quantum} 
that ``..quantum fluctuations may be of electromagnetic origin..'' This 
is to be understood in the sense that the stochastic mechanism, which 
underlies quantum mechanics, could be due to the ``non-deterministic'' 
behavior of charged particles, as a consequence of interaction with 
their own field~\cite{nelson:quantum}. This conjecture is - using a 
different language - very similar to the above. Finally we note, that 
there is a certain overlap of the present ideas with those underlying
Stochastic Electrodynamics~\cite{pena.cetto:stochastic}.

\section{Concluding remarks}
\label{sec:7}
In this work Schr\"odinger's equation with gauge coupling has been 
derived. The construction was based on the validity of a continuity
equation and a linear differential equation for a complex-valued 
state variable. As an heuristic recipe this scheme may  probably be 
extended to more general situations. Obvious possibilities to do 
that include a relativistic formulation or a multi-component version, 
which should be the analogon of non-abelian gauge theories. Other types 
of variables (quaternions) or even an application beyond the 
realm of special relativity seem conceivable. From a physical point 
of view, the most important generalization or modification of the 
present theory concerns the assumption of a complex state variable. 
This is a purely mathematical assumption which  - despite of the outstanding 
structural properties of the field of complex numbers - should be 
replaced by an different, equivalent assumption which can be directly 
interpreted  in physical terms. This is the most challenging 
question to be asked in the context of the present theory. 
\begin{appendix}
\section{Calculation of $\chi$ and $F$}
\label{sec:8}  
Combining the derivative of ~(\ref{eq:BCOND5}) with respect to
$\rho$ with ~(\ref{eq:BCOND2}), one obtains
\begin{equation}
  \label{eq:SZWR1}
  \frac{\partial\bar{d}_1}{\partial \rho}\frac{\partial\chi_1}{\partial \rho}- 
\frac{\partial\bar{d}_2}{\partial \rho}\frac{\partial\chi_2}{\partial \rho}=0
\mbox{.}
\end{equation}
Using~(\ref{eq:SZWR1}) and~(\ref{eq:BCOND5}) one obtains 
an elementary differential equation, which shows that 
the ratio  of $\bar{d}_1$ and $\bar{d}_2$ does not 
depend on $\rho$; it is convenient to write this in the form       
\begin{equation}
  \label{eq:SZWR2}
\frac{\bar{d}_1}{\bar{d}_2}=\mathrm{e}^{f(S)}
\mbox{,}
\end{equation}
where $f(S)$ is an unknown function. 
If~(\ref{eq:BCOND7}) is multiplied by $2\rho$ and afterwards 
combined with ~(\ref{eq:BCOND8}), one obtains a differential 
equation 
\begin{equation}
  \label{eq:SZWR3}
2\rho\frac{\partial U}{\partial \rho}=U
\mbox{}
\end{equation}
for the $\rho-$dependence of a quantity $U(\rho,S)$, defined by
\begin{equation}
  \label{eq:SZWR4}
U(\rho,S)=\frac{\bar{d}_1}{\bar{d}_2} \frac{\partial\chi_1}{\partial S}-\frac{\partial\chi_2}{\partial S}
\mbox{.}
\end{equation}
Inserting the solution of~(\ref{eq:SZWR3}) 
in~(\ref{eq:BCOND8}) leads to the important intermediate
result
\begin{eqnarray}
\bar{d}_2(\rho,S)&=& \frac{\sqrt{\rho}}{m}\mathrm{e}^{-h(S)}
  \label{eq:WIZWI1}
\mbox{,} \\
\bar{d}_1(\rho,S)&=& \frac{\sqrt{\rho}}{m}\mathrm{e}^{-h(S)+f(S)} 
\label{eq:WIZWI2}
\mbox{.} 
\end{eqnarray}  
Thus, the dependence of $\bar{d}_1$ and $\bar{d}_2$ on $\rho$ is
known while the two functions $h$ and $f$, which both depend only 
on $S$, are still to be found.

The solution of~(\ref{eq:BCOND5}) and~(\ref{eq:BCOND6}) with 
regard to $\frac{\partial\chi_1}{\partial \rho}$ and $\frac{\partial\chi_2}{\partial \rho}$ is given by 
\begin{equation}
  \label{eq:WIJWI3}
\frac{\partial\chi_1}{\partial \rho}=\frac{\bar{d}_2}{\bar{d}_2 \bar{a}_1-\bar{d}_1 \bar{a}_2},
\hspace{0.7cm}
\frac{\partial\chi_2}{\partial \rho}=\frac{\bar{d}_1}{\bar{d}_2 \bar{a}_1-\bar{d}_1 \bar{a}_2}
\mbox{.}
\end{equation}
The $\bar{a}_i$ may be expressed, using~(\ref{eq:DEF1}) 
and~(\ref{eq:DEF3}), by the $\bar{d}_i$; this determines the 
$\rho$-dependence of the right hand sides of the differential 
equations~(\ref{eq:WIJWI3}). Integrating~(\ref{eq:WIJWI3}) 
yields
\begin{eqnarray}
\chi_1(\rho,S)&=& -\frac{2m|d|^2}{c_2}\sqrt{\rho}
\frac{\mathrm{e}^{h(S)-f(S)}}{\mathrm{e}^{f(S)}+
\mathrm{e}^{-f(S)}} + G(S)
  \label{eq:DIGGL1}
\mbox{,} \\
\chi_2(\rho,S)&=& -\frac{2m|d|^2}{c_2}\sqrt{\rho}
\frac{\mathrm{e}^{h(S)}}{\mathrm{e}^{f(S)}+
\mathrm{e}^{-f(S)}} + H(S)
\label{eq:DIGGL2}
\mbox{,} 
\end{eqnarray}
where $|d|^2=d_1^2+d_2^2$ and $G(S)$, $H(S)$ are unknown functions of 
$S$. In~(\ref{eq:DIGGL1}) and~(\ref{eq:DIGGL2}) 
the abbreviations 
\begin{equation}
  \label{eq:ABKKN}
c_1=a_1d_1+a_2d_2,
\hspace{0.7cm}
c_2=a_2d_1-a_1d_2,
\mbox{}
\end{equation}
have been used. The remaining task is the calculation of the four
functions $f,h,G,H$.

In order to do that we insert the solution 
of~(\ref{eq:SZWR3}) for $U$ in~(\ref{eq:SZWR4}) and 
combine it with~(\ref{eq:SZWR2}). This leads to the relation        
\begin{equation}
  \label{eq:HZAWE1}
\mathrm{e}^{f(S)}\frac{\partial\chi_1}{\partial S}-\frac{\partial\chi_2}{\partial S}=
\mathrm{e}^{h(S)}\sqrt{\rho}
\mbox{.}
\end{equation}
Using~(\ref{eq:HZAWE1}) and~(\ref{eq:BCOND1}) we obtain the 
following derivatives of $\chi_1,\;\chi_2$ with respect to $S$.   
\begin{eqnarray}
\frac{\partial\chi_1}{\partial S}&=&\sqrt{\rho}\frac{c_2\mathrm{e}^{f(S)}+c_1}
{c_2\mathrm{e}^{2f(S)}+c_2}\mathrm{e}^{h(S)}
\label{eq:ABCH12}
\mbox{,}\\
\frac{\partial\chi_2}{\partial S}&=&\sqrt{\rho}\frac{c_1\mathrm{e}^{f(S)}-c_2}
{c_2\mathrm{e}^{2f(S)}+c_2}\mathrm{e}^{h(S)}
\label{eq:ABCH22}
\mbox{.}
\end{eqnarray}
Next, we calculate the derivatives of $\chi_1$ und $\chi_2$ 
[see ~(\ref{eq:DIGGL1}) und~(\ref{eq:DIGGL2})] with 
respect to $S$ and equate the results with the derivatives 
as given by ~(\ref{eq:ABCH12}) und~(\ref{eq:ABCH22}).
From the resulting relation we may draw two conclusions. The 
first implies that $G(S)$ and $H(S)$ must be constants, say
\begin{equation}
  \label{eq:AKGLJ1}
G(S)=C_3 \mbox{,}
\hspace{0.5cm}
H(S)=C_4
\mbox{.}
\end{equation} 

The second implies two differential equations for 
$f(S)$ and $h(S)$, namely
\begin{eqnarray}
-\frac{2m|d|^2}{c_2}\frac{\partial}{\partial S}\frac{1}{L_1(S)}&=&
\frac{c_2\mathrm{e}^{f(S)}+c_1}
{c_2L_1(S)}
\label{eq:GLFFH1}
\mbox{,}\\
-\frac{2m|d|^2}{c_2}\frac{\partial}{\partial S}\frac{1}{L_2(S)}&=&
\frac{c_1\mathrm{e}^{f(S)}-c_2}
{c_2L_1(S)}
\label{eq:GLFFH2}
\mbox{.}
\end{eqnarray} 
Here, the abbreviations
\begin{equation*}
L_1(S)= \mathrm{e}^{-h(S)}\left(1+
\mathrm{e}^{2f(S)} \right)\mbox{,}
\hspace{0.5cm}
L_2(S)= \mathrm{e}^{-f(S)} L_1(S)
\mbox{}
\end{equation*}     
have been used. After a short 
rearrangement, Eqs.~(\ref{eq:GLFFH1}),
~(\ref{eq:GLFFH2}) take the form
\begin{eqnarray}
-\frac{\partial h}{\partial S}+\frac{\partial f}{\partial S}\frac{2\mathrm{e}^{2f}}
{1+\mathrm{e}^{2f}}&=&\frac{1}{2m|d|^2}
\left(c_2\mathrm{e}^{f}+c_1 \right)
\label{eq:EKUDG1}
\mbox{,}\\
\frac{\partial h}{\partial S}-\frac{\partial f}{\partial S}\frac{\mathrm{e}^{f}-\mathrm{e}^{-f}}
{\mathrm{e}^{f}+\mathrm{e}^{-f}}&=&\frac{1}{2m|d|^2}
\left(c_2\mathrm{e}^{-f}-c_1 \right)
\label{eq:EKUDG2}
\mbox{,}
\end{eqnarray}  
which shows, that the two equations may be decoupled. Addition 
of Eqs.~(\ref{eq:EKUDG1}) and~(\ref{eq:EKUDG2}) yields a
differential equation for $f(S)$ alone, namely
\begin{equation}
  \label{eq:EDGLFA}
\frac{\partial f}{\partial S}=\frac{c_2}{2m|d|^2}
\left(\mathrm{e}^{f}+\mathrm{e}^{-f} \right)
\mbox{,}
\end{equation}
which may be solved easily. From~(\ref{eq:EDGLFA}) 
and~(\ref{eq:EKUDG1}) we obtain for the functions 
$f(S)$ and $h(S)$ the final results 
\begin{eqnarray}
f(S)&=&\ln \tan \left(\frac{c_2}{2m|d|^2}S+C_5 \right)
\label{eq:DGRFF}
\mbox{,}\\
h(S)&=&-\ln \cos \left(\frac{c_2}{2m|d|^2}S+C_5 \right)-
\frac{c_1}{2m|d|^2}S-C_6
\label{eq:DGRFH}
\mbox{,}
\end{eqnarray}  
containing the constants of integration $C_5,\,C_6$ 
and the coefficients $c_i,\,d_i$. 

This determines the functional form of the variables $\chi$ 
and $F$, we were looking for. Using Eqs.~(\ref{eq:DIGGL1}),~(\ref{eq:DIGGL2}),~(\ref{eq:DGRFF}), and~(\ref{eq:DGRFH}) one 
obtains~(\ref{eq:DRFXI1}) and ~(\ref{eq:DRFXI2}) for the real- and 
imaginary parts of $\chi$. For the real- and imaginary 
parts of $F$ one obtains with the help of~(\ref{eq:DEF3}),~(\ref{eq:WIZWI1}), 
and~(\ref{eq:WIZWI2}) the expressions ~(\ref{eq:DRFFF1}) 
and ~(\ref{eq:DRFFF2}).

\section{Calculation of parameters}
\label{sec:9}  

Conditions~(\ref{eq:BCOND3NE}),~(\ref{eq:BCOND4ANE})~(\ref{eq:BCOND4BNE})
and~(\ref{eq:BCOND9NE}) have not yet been used and play the role of 
constraints for our parameters. In this appendix we  
insert $\chi$ and $F$ [see 
Eqs.~(\ref{eq:DRFXI1NE}),~(\ref{eq:DRFXI2NE}),~(\ref{eq:DRFFF1}), 
and ~(\ref{eq:DRFFF2})] 
in ~(\ref{eq:BCOND3NE}),~(\ref{eq:BCOND4ANE})~(\ref{eq:BCOND4BNE})
and~(\ref{eq:BCOND9NE}) and evaluate the resulting parameters.

We introduce the following abbreviations:
\begin{eqnarray}
U&=&\frac{c_2}{2m|d|^2}S+C_5
\label{eq:ABKUE1}
\mbox{,} \\
V&=&\frac{c_1}{2m|d|^2}S+C_6 
 \label{eq:ABKUE2}
\mbox{,} \\
W&=& \frac{2m|d|^2}{c_2}\sqrt{\rho}
\label{eq:ABKUE3}
\mbox{,} \\
T&=&\frac{\sqrt{\rho}}{m|d|^2} 
\label{eq:ABKUE4}
\mbox{.}
\end{eqnarray} 
Using these abbreviations, the real and imaginary parts of $\chi$ and $F$ 
are written as
\begin{eqnarray}
\chi_1&=&-W \cos U \mathrm{e}^{-V}+C_3 
\label{eq:REXABKUE}
\mbox{,} \\
\chi_2&=&-W \sin U \mathrm{e}^{-V}+C_4 
 \label{eq:IMXABKUE}
\mbox{,}\\
F_1&=&T\mathrm{e}^{V}\left(d_1 \sin U + d_2 \cos U \right)
\label{eq:REFABKUE}
\mbox{,} \\
F_2&=&T\mathrm{e}^{V}\left(d_1 \cos U - d_2 \sin U \right)
 \label{eq:IMFABKUE}
\mbox{,}
\end{eqnarray}
and the functions $\bar{a}_i,\,\bar{b}_i,\,\bar{d}_i,\,\bar{e}_i,\;i=1,2$
[see~(\ref{eq:DEF1})-~(\ref{eq:DEF4})] take the form
\begin{eqnarray}
\bar{a}_1&=& T\mathrm{e}^{V}
\left(c_1 \sin U - c_2 \cos U \right)
\label{eq:A1ABKUE}
\mbox{,} \\
\bar{a}_2&=& T\mathrm{e}^{V}
\left(c_1 \cos U + c_2 \sin U \right)
 \label{eq:A2ABKUE}
\mbox{,}\\
\bar{b}_1&=& T\mathrm{e}^{V}
\left(g_1 \sin U - g_2 \cos U \right)
\label{eq:B1ABKUE}
\mbox{,} \\
\bar{b}_2&=& T\mathrm{e}^{V}
\left(g_1 \cos U + g_2 \sin U \right)
 \label{eq:B2ABKUE}
\mbox{,} \\
\bar{d}_1&=& |d|^{2}\,T \mathrm{e}^{V} \sin U
\label{eq:D1ABKUE}
\mbox{,} \\
\bar{d}_2&=& |d|^{2}\, T \mathrm{e}^{V} \cos U
 \label{eq:D2ABKUE}
\mbox{,}\\
\bar{e}_1&=& T\mathrm{e}^{V}
\left(h_1 \sin U - h_2 \cos U \right)
\label{eq:E1ABKUE}
\mbox{,} \\
\bar{e}_2&=& T\mathrm{e}^{V}
\left(h_1 \cos U + h_2 \sin U \right)
 \label{eq:E2ABKUE}
\mbox{,}
\end{eqnarray}
where $g_i,\,h_i$ are defined by 
\begin{align}
&g_1=b_1d_1+b_2d_2,\;\;g_2=b_2d_1-b_1d_2\mbox{,}\label{eq:NGDEF}\\
&h_1=e_1d_1+e_2d_2,\;\;h_2=e_2d_1-e_1d_2\mbox{.}\label{eq:NHDEF}
\end{align} 
Inserting $\chi_i,\,F_i$ in~(\ref{eq:BCOND3NE}) we obtain 
\begin{equation}
  \label{eq:BEDCNL}
c_1=0
\mbox{.}
\end{equation}
The latter relation occurred already in section~\ref{sec:3}); it implies  
that the variable $S$ appears only in the phase and not in the 
modulus of $\chi$. From the conditions~(\ref{eq:BCOND4ANE}) 
and~(\ref{eq:BCOND4BNE}) we obtain the relations 
\begin{eqnarray}
g_2&+&2
|d|^{2}\left(
S\frac{\partial\tilde{u}}{\partial q}+ \frac{\partial C_5}{\partial q}
 \right)=0
\label{eq:BC4ABD}
\mbox{,} \\
g_1 &-& 2|d|^{2}\frac{\partial C_6}{\partial q}=0 
\label{eq:BC4BBD}
\mbox{,} 
\end{eqnarray}
where the abbreviation
\begin{equation}
  \label{eq:ABKUSCHL}
\tilde{u}=\frac{c_2}{2m|d|^{2}}
\mbox{,}
\end{equation}
has been used. Eq.~(\ref{eq:BEDCNL}), i.e. $V=C_6$, has already 
been used in~(\ref{eq:BC4BBD}), and will also be used in the 
following to simplify the formulas. Since Eq.~(\ref{eq:BC4ABD}) must 
hold for arbitrary $S$, condition~(\ref{eq:BC4ABD}) implies the  
two constraints 
\begin{equation}
  \label{eq:BC4ABDAA}
\frac{\partial \tilde{u}}{\partial q} = 0
\mbox{,}
\end{equation}
($\tilde{u}$ depends only on $t$) and 
\begin{equation}
  \label{eq:BC4ABDBB}
g_2+2|d|^{2} \frac{\partial C_5}{\partial q} = 0
\mbox{.}
\end{equation}

The remaining condition~(\ref{eq:BCOND9NE}) leads to a lenghty 
expression. We rewrite condition~(\ref{eq:BCOND9NE}) in the following
form 
\begin{eqnarray}
T_e&+&T_a+T_b+T_d = 0
\label{eq:UDLTDW}
\mbox{,} \\
T_e&=&\bar{e_1}\chi_1-\bar{e_2}\chi_2
\label{eq:UDLTDWTE}
\mbox{,}\\
T_a&=&\bar{a_1}\frac{\partial\chi_1}{\partial t}-\bar{a_2}\frac{\partial\chi_2}{\partial t}
\label{eq:UDLTDWTA}
\mbox{,}\\
T_b&=&\bar{b_1}\frac{\partial\chi_1}{\partial q}-\bar{b_2}\frac{\partial\chi_2}{\partial q}
\label{eq:UDLTDWTB}
\mbox{,}\\
T_d&=&\bar{d_1}\frac{\partial^{2} \chi_1}{\partial q^{2}}-\bar{d_2}\frac{\partial^{2} \chi_2}{\partial q^{2}}
\label{eq:UDLTDWTD}
\mbox{,}
\end{eqnarray}
associating a term with each one  of the coefficients $e,\,a,\,b,\,d$.
Inserting $\chi_i,\,F_i$ we obtain (using $c_1=0$) 
\begin{eqnarray}
T_e&=&T W h_2 + \\
& &T\mathrm{e}^{C_6}
\big[
\left(C_3h_1-C_4h_2\right)\sin U-
\left(C_4h_1+C_3h_2\right)\cos U
\big]
\label{eq:AQETDWTE}
\mbox{,}\\
T_a&=&-T\frac{\sqrt{\rho}}{\tilde{u}} c_2
\left( 
\frac{1}{\tilde{u}}\frac{\partial\tilde{u}}{\partial t}+\frac{\partial C_6}{\partial t}
\right)+\\
& &T\mathrm{e}^{C_6}
\bigg[
-c_2\frac{\partial C_4}{\partial t}\sin U-c_2\frac{\partial C_3}{\partial t}\cos U
\bigg]
\label{eq:AQETDWTA}
\mbox{,}\\
T_b&=&T\frac{\sqrt{\rho}}{\tilde{u}}
\bigg[
g_1\left(S\frac{\partial\tilde{u}}{\partial q}+\frac{\partial C_5}{\partial q} \right)-
g_2\left(\frac{1}{\tilde{u}}\frac{\partial\tilde{u}}{\partial q}+\frac{\partial C_6}{\partial q} \right)
\bigg]+\\
& &T\mathrm{e}^{C_6}
\bigg[
\left(g_1\frac{\partial C_3}{\partial q} -g_2\frac{\partial C_4}{\partial q}\right)\sin U-
\left(g_2\frac{\partial C_3}{\partial q} +g_1\frac{\partial C_4}{\partial q}\right)\cos U
\bigg]
\label{eq:AQETDWTB}
\mbox{,}\\
T_d&=&-T\frac{\sqrt{\rho}}{\tilde{u}}|d|^{2} 
\bigg[
2\left(\frac{1}{\tilde{u}}\frac{\partial\tilde{u}}{\partial q}+\frac{\partial C_6}{\partial q} \right)
\left(S\frac{\partial\tilde{u}}{\partial q}+\frac{\partial C_5}{\partial q} \right)-\\
& &\left(S\frac{\partial^{2}\tilde{u}}{\partial q^{2}}+\frac{\partial^{2}C_5}{\partial q^{2}} \right)
\bigg]+\\
& &T \mathrm{e}^{C_6}|d|^{2}
\bigg[
\frac{\partial^{2} C_3}{\partial q^{2}}\sin U-\frac{\partial^{2} C_4}{\partial q^{2}}\cos U
\bigg]
\label{eq:AQETDWTD}
\mbox{.}
\end{eqnarray}
The left hand side of Eq.~(\ref{eq:UDLTDW}) is a sum of four
terms, each one beeing proportional to a 
linear independent function of $\rho,S$, namely
$\rho,\,\rho S,\,\sin S,\, \cos S$. As a consequence 
Eq.~(\ref{eq:UDLTDW}) leads to maximal four sub-conditions. 
The vanishing of the coefficient of $\rho$ implies
\begin{equation}
\label{eq:AWEGGBA1}
\begin{split}
&2\frac{h_2}{c_2}
-2 \left( \frac{1}{\tilde{u}}\frac{\partial\tilde{u}}{\partial t}+\frac{\partial C_6}{\partial t} \right)
+ \frac{2}{c_2}
\left[
g_1 \frac{\partial C_5}{\partial q}
-g_2\left( \frac{1}{\tilde{u}}\frac{\partial\tilde{u}}{\partial q}+\frac{\partial C_6}{\partial q} \right)
 \right] \\ 
&-2 \frac{|d|^{2}}{c_2}
\left[
2\frac{\partial C_5}{\partial q}
\left( \frac{1}{\tilde{u}}\frac{\partial\tilde{u}}{\partial q}+\frac{\partial C_6}{\partial q} \right)
-\frac{\partial^{2}C_5}{\partial q^{2}}
 \right]=0
\end{split}
\mbox{,}
\end{equation}
The condition that the coefficient of $\rho S$ vanishes does not lead to a 
new constraint; it is automatically fulfilled if Eq.~(\ref{eq:BC4ABDAA}) 
holds. Finally, the condition of a vanishing coefficient of 
$\sin S,\, \cos S$ leads to the two relations
\begin{eqnarray}
C_3 h_1-C_4 h_2 - c_2 \frac{\partial C_4}{\partial t}+
\left(g_1\frac{\partial C_3}{\partial q} - g_2\frac{\partial C_4}{\partial q} \right)
+|d|^{2}\frac{\partial^{2} C_3}{\partial q^{2}}&=&0
\label{eq:ANHZGGBC1}
\mbox{,} \\ 
-C_4 h_1-C_3 h_2 - c_2 \frac{\partial C_3}{\partial t}-
\left(g_2\frac{\partial C_3}{\partial q} + g_1\frac{\partial C_4}{\partial q} \right)
-|d|^{2}\frac{\partial^{2} C_4}{\partial q^{2}}&=&0
\label{eq:ANHZGGBC2}
\mbox{,}
\end{eqnarray}
Now all constraints for the 12 real functions  
$a_i$, $b_i$,$d_i$, $e_i$, $C_3$, $C_4$, $C_5$, $C_6$ 
are known explicitely. These are Eqs.~(\ref{eq:AWEGGBA1})-(\ref{eq:ANHZGGBC2})
and the above 
relations~(\ref{eq:BEDCNL}),~(\ref{eq:BC4ABDAA}),~(\ref{eq:BC4BBD}),~(\ref{eq:BC4ABDBB}). Using~(\ref{eq:BC4ABDAA}), (\ref{eq:BC4BBD}) 
and~(\ref{eq:BC4ABDBB}), Eq.~(\ref{eq:AWEGGBA1}) takes the form
\begin{equation}
\label{eq:AWEJJUK1}
2\frac{h_2}{c_2}
-2 \left( \frac{1}{\tilde{u}}\frac{\partial\tilde{u}}{\partial t}+\frac{\partial C_6}{\partial t} \right)
- \frac{g_1 g_2}{c_2 |d|^{2}} \\
+2 \frac{|d|^{2}}{c_2} \frac{\partial^{2}C_5}{\partial q^{2}} =0
\mbox{.}
\end{equation}  
With the help of~(\ref{eq:BC4BBD}),~(\ref{eq:BC4ABDBB}) one obtains 
\begin{equation}
\label{eq:AWNAFRK2}
h_2 =
2 m |d|^{2} \tilde{u} 
\left( \frac{1}{\tilde{u}}\frac{\partial\tilde{u}}{\partial t}+\frac{\partial C_6}{\partial t} \right)
- 2 |d|^{2} \frac{\partial C_5}{\partial q} \frac{\partial C_6}{\partial q}
-|d|^{2} \frac{\partial^{2}C_5}{\partial q^{2}} 
\mbox{.}
\end{equation}
The latter formula shows that $h_2$ is proportional to $|d|^{2}$, where
the constant of proportionality is determined by the functions 
$\tilde{u},\,C_5$ and $C_6$. Exactly the same holds for 
$c_1,\,c_2,\,g_1,\,g_2$. The definitions of $c_i,\,g_i,\,h_i$ show 
that it should be possible, for given $c_i,\,g_i,\,h_i$, to express the 
$a_i,\,b_i,\,e_i$ in terms of the $d_i$. The resulting 
relations (for the complex quantities $a,\,b,\,e$) 
are given by Eqs.~(\ref{eq:FJKUZA})-(\ref{eq:FJKUZE}).
The coefficients on the left hand side of Eq.~(\ref{eq:ANHZGGBC1}) 
and~(\ref{eq:ANHZGGBC2}) are arbitrary functions. This implies the
(expected) trivial solutions
\begin{equation}
  \label{eq:EEGFFM4}
C_3=0,\;\;\;C_4=0
\mbox{}
\end{equation}
for these equations. As a consequence, the constant $d$ drops out of the 
equations, the remaining arbitrary parameters beeing 
$C_5,\,C_6,\,H_1,\,\tilde{u}$.  

\end{appendix}

\bibliography{uftbig}
\end{document}